\documentclass[10pt,journal,compsoc]{IEEEtran}

\RequirePackage[abbreviate=true, dateabbrev=true, natbib=true, isbn=false, doi=false, urldate=comp, url=false, maxbibnames=9, maxcitenames=1, backref=false, backend=biber, style=ACM-Reference-Format, language=american,firstinits=true]{biblatex}
\usepackage{microtype}
\usepackage{booktabs} % For formal tables
\usepackage{arydshln} % for dashed lines in tables (tabu is much more flexible)
\usepackage{graphicx}
\usepackage{wrapfig}
\graphicspath{ {images/} }
\usepackage{pdfpages}
\usepackage{blindtext}
\usepackage{subfigure}
\usepackage{caption}

\usepackage{amsmath}
\allowdisplaybreaks
\usepackage{bm} % produces better bold symbols than does \boldsymbol{x}; use \bm{x}
\usepackage{siunitx}
\usepackage{amsmath, mathrsfs,amsfonts}
\usepackage{wasysym} % smiley face
\usepackage{xfrac} % use command \sfrac{1}{2} to get a nice 1/2 symbol.

\usepackage{savesym} % openbox symbol conflict between tx fonts and amsthm
\savesymbol{openbox}
\usepackage{cleveref} % easier references for sections, figures etc.
\usepackage{commath}
\usepackage[export]{adjustbox}

\usepackage{float}
\usepackage{array}
\newcolumntype{L}{>{\arraybackslash}m{7cm}}
\usepackage{multirow}

\usepackage{stfloats}
\usepackage{url}
\usepackage{glossaries}
\newacronym{iot}{IoT}{Internet of Things}
\newacronym{gps}{GPS}{Global Positioning System}
\newacronym{cbprp}{C-BPRP}{Cooperative Bayesian Packet Reception Probability}
\newacronym{bprp}{B-PRP}{Bayesian Packet Reception Probability}
\newacronym{rssi}{RSSI}{Receiver Signal Strength Indicator}
\newacronym{ble}{BLE}{Bluetooth Low Energy}
\newacronym{aoa}{AoA}{Angle of Arrival}
\newacronym{tof}{ToF}{Time of Flight}
\newacronym{nlos}{NLoS}{Non-Line of Sight}
\newacronym{los}{LoS}{Line of Sight}
\newacronym{mcmc}{MCMC}{Markov Chain Mote-Carlo}
\newacronym{saf}{SAF}{Stack Attenuation Factor}
\usepackage{tikz}
\usetikzlibrary{shapes,arrows,patterns}

\usepackage{enumitem}
\setlist{left=0pt .. \parindent}
\begin{document}
% \sloppy
% \title{\large{Finding by Counting: A Probabilistic Packet Count Model for Indoor Localization in BLE Environments}}
% \title{Packet Reception Probability: A robust parameter for scalable, commodity-hardware based indoor localization}
% \title{The Packet You Can’t Decode Will Help Find You: Packet Reception Probability for a Robust, Scalable, Commodity-Hardware Based Indoor Localization}
% \title{Extracting Location from Packets That You Can't Decode}
% \title{The Packet You Can’t Decode Will Help Locate You}
% \title{Your Location from Packets That You Can't Decode}
\title{Packet Reception Probability: Packets That You Can't Decode Can Help Keep You Safe}

% \begin{abstract}
%   \input{abstract-hs}
% \end{abstract}
% <concept_id>10003120.10003138.10003140</concept_id>
% <concept_desc>Human-centered computing~Ubiquitous and mobile computing systems and tools</concept_desc>
% \ccsdesc[500]{Human-centered computing~Ubiquitous and mobile computing systems and tools}

%\tile{Improving the accuracy of localization in IoT BLE spaces through robust packet perception model}
%\titlenote{Produces the permission block, and
%  copyright information}
%\subtitle{Extended Abstract}
%\subtitlenote{The full version of the author's guide is available as
%  \texttt{acmart.pdf} document}

%\subtitle{[Anonymous Authors]}
% \author{Anonymous Authors}
% \author{Subham~De,~\IEEEmembership{University of Illinois at Urbana-Champaign,}
%         John~Doe,~\IEEEmembership{University of Illinois at Urbana-Champaign,}
%         and~Jane~Doe,~\IEEEmembership{University of Illinois at Urbana-Champaign}}

\author{\IEEEauthorblockN{Subham De\IEEEauthorrefmark{1},
Deepak Vasisht\IEEEauthorrefmark{2}, Hari Sundaram\IEEEauthorrefmark{3} and
Robin Kravets\IEEEauthorrefmark{4}}\\
\IEEEauthorblockA{Computer Science,
University of Illinois at Urbana-Champaign\\
% Urbana, Illinois\\
Email: \IEEEauthorrefmark{1}de5@illinois.edu,
\IEEEauthorrefmark{2}deepakv@illinois.edu,
\IEEEauthorrefmark{3}hs1@illinois.edu,
\IEEEauthorrefmark{4}rhk@illinois.edu}}

\IEEEtitleabstractindextext{%
\begin{abstract}
This paper provides a robust, scalable Bluetooth Low-Energy (BLE) based indoor localization solution using commodity hardware. While WiFi-based indoor localization has been widely studied, BLE has emerged a key technology for contact-tracing in the current pandemic. To accurately estimate distance using BLE on commercial devices, systems today rely on Receiver Signal Strength Indicator(RSSI) which suffers from sampling bias and multipath effects. We propose a new metric: Packet Reception Probability (PRP) that builds on a counter-intuitive idea that we can exploit packet loss to estimate distance. We localize using a Bayesian-PRP formulation that also incorporates an explicit model of the multipath. To make deployment easy, we do not require any hardware, firmware, or driver-level changes to off-the-shelf devices, and require minimal training. PRP can achieve meter level accuracy with just 6 devices with known locations and 12 training locations. We show that fusing PRP with RSSI is beneficial at short distances (\SI[]{\le 2}{\meter}). Beyond \SI[]{\ge 2}{\meter}, fusion is worse than PRP, as RSSI becomes effectively de-correlated with distance. Robust location accuracy at all distances and ease of deployment with PRP can help enable wide range indoor localization solutions using BLE. 
% This paper provides a robust, scalable Bluetooth Low-Energy (BLE) based indoor localization solution using commodity hardware. While WiFi-based indoor localization has been widely studied, BLE has emerged a key technology for contact-tracing in the current pandemic. To accurately estimate distance using BLE, we need to overcome key challenges in Receiver Signal Strength Indicator (RSSI), including sampling bias and multipath effects. We propose a new metric: Packet Reception Probability (PRP) that builds on a counter-intuitive idea that we can exploit packet loss to estimate distance. We localize using a Bayesian-PRP formulation.  B-PRP encodes an explicit model of the multipath and produces significant improvements over RSSI methods in two public spaces. B-PRP can achieve meter level accuracy with just 6 beacon locations known and 12 training locations.
% % We show that sampling bias severely degrades RSSI based distance estimates beyond \SI[]{\ge 2}{m}.
% We show that fusing B-PRP with RSSI is beneficial at short distances (\SI[]{\le 2}{\meter}). Beyond \SI[]{\ge 2}{\meter}, fusion is worse than B-PRP, as RSSI becomes effectively de-correlated with distance. Our solution does not require any hardware, firmware, or driver-level changes to off-the-shelf devices, and involves minimal deployment and re-training costs. We also design a triangle inequality based joint estimation problem that helps us to directly estimate contact tracing distance between two individuals rather than estimating their locations first, which gives us 10\% performance improvement.

\end{abstract}}

\maketitle
\section{introduction}
\label{sec:introduction}

Indoor positioning is a widely studied problem in academia and industry~\cite{vasisht2016decimeter,vasisht2018duet,kotaru2015spotfi, liu2017smartlight, youssef2005horus,he2016wi,tian2018improve}. Coupled with the high penetration of consumer radio devices (e.g. smartphones), indoor positioning can re-imagine use of indoor spaces like retail spaces, malls, museums, and warehouses. Today, the contact-tracing challenge due to the pandemic has put an urgent, renewed focus on developing a robust, low-cost, scalable, indoor localization solution. Indoor-localization based contact-tracing\footnote{Contact-tracing requires us to calculate relative distance between individuals. Inferring relative distance from location is straightforward.} that helps us determine if a pair of individuals are ``social-distancing,'' separated by more than 6ft, may safely re-open the world economy.

Technological solutions for contact tracing that use smartphones are an important complement to normative (e.g., wearing a mask) and policy (e.g. stay-at-home) interventions for mitigating effects of the pandemic. Bluetooth Low-Energy (BLE) is emerging as the key contact-tracing technology and is being used in contact-tracing apps around the world. For example, the Aarogya Setu contact-tracing app\footnote{\url{https://www.mygov.in/aarogya-setu-app/}} in India, uses BLE and has been downloaded 120M times. The open-source, privacy-preserving contact-tracing framework,  BlueTrace\footnote{\url{https://bluetrace.io}} (deployed in Singapore) uses BLE packets to detect presence (i.e., a smartphone that can hear another must be in proximity of the other.), \textit{not} distance. BLE is preferable to WiFi for contact-tracing: BLE uses $10\times$ \textit{less power} than does WiFi; BLE can be easily used to infer the presence of nearby peers without presence of WiFi infrastructure. The newly proposed Exposure Notification Service by Apple-Google\footnote{\url{https://www.apple.com/covid19/contacttracing/}} also relies on BLE beacons and signal strength measurements.

% Bluetooth Low-Energy based apps for contact-tracing have two well-known shortcomings. These apps primarily use either RSSI (Received Signal Strength Indicator) or presence to determine risk to COVID exposure . Prior work~\cite{bahl2000radar, heurtefeux2012rssi, zanella2016best} demonstrates that RSSI-based methods experience large errors (order of several meters) in positioning, especially in the low RSSI-large distance regime. RSSI has an important benefit: it is present on all modern devices.  In contrast, we cannot use CSI (Channel State Information)~\cite{kotaru2015spotfi, xiong2013arraytrack, vasisht2016decimeter}, a recent method that enables sub-meter accuracy since off-the-shelf devices typically do not report CSI. A recent work ~\cite{schulz2018shadow} has enabled CSI for WiFi in some smartphones, but cannot be applied for BLE. Some contact-tracing apps also use `presence'---if one device can hear another---to determine if an individuals is close to another infected person. Presence is a poor proxy for distance since devices can hear Bluetooth beacons well beyond 6 ft social distancing radius, and also hear them across aisles and walls.

\subsection{Overcoming Key Technological Limitations for Contact Tracing}
\label{sub:Overcoming Key Limitations}

\begin{figure*}[t]
  %\begin{subfigure}[t]{0.7\columnwidth}
  \centering
  \subfigure[]{\label{fig:packetcountIntroduction_sub1}\includegraphics[viewport=0.1in 0.3in 5.8in 5in,width=0.6\columnwidth, clip=true]{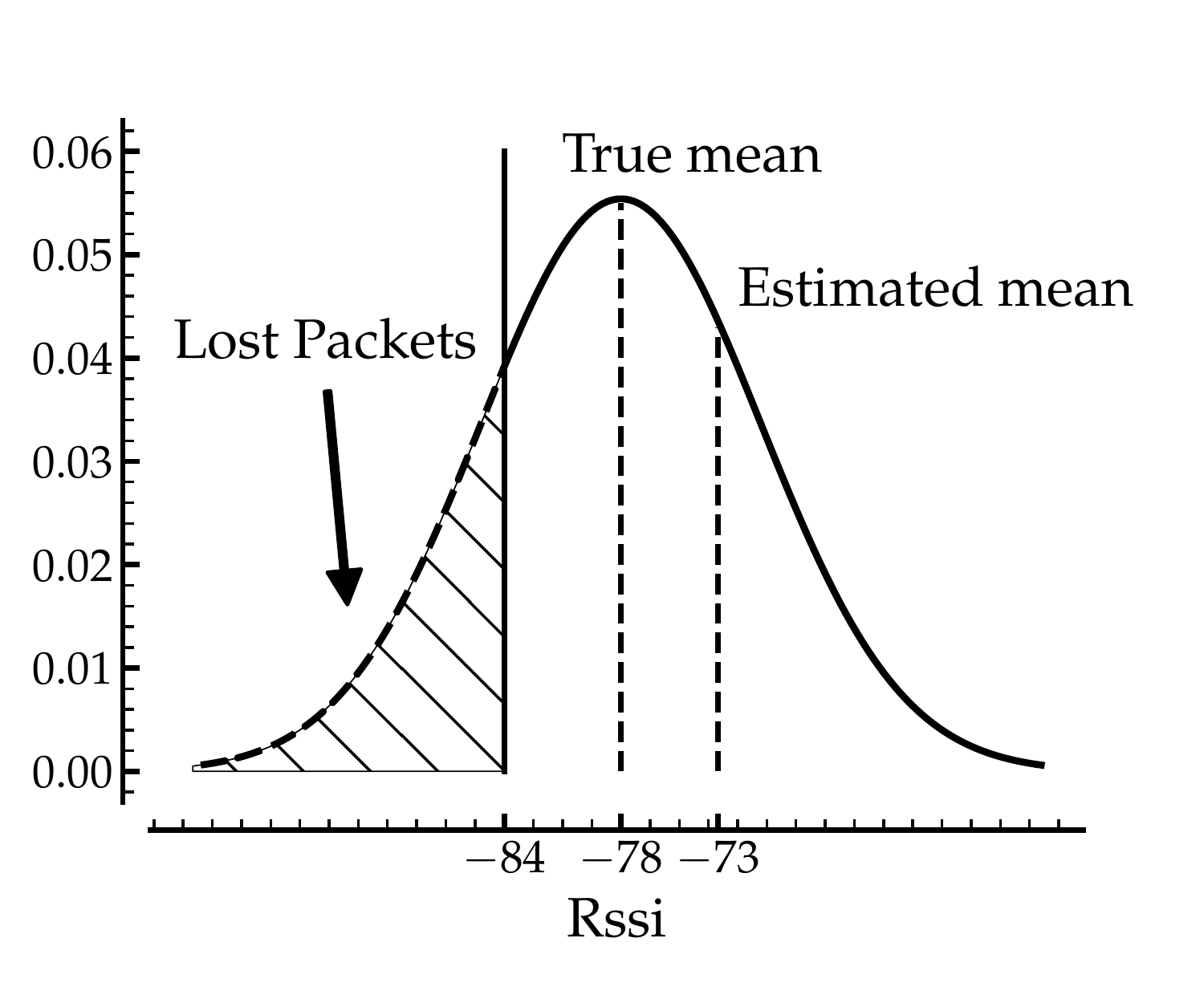}}\quad
  %  \subfigure[]{\label{fig:packetcountIntroduction_sub2}\includegraphics[viewport=0.9in 0.6in 4.2in 3.9in,width=0.45\columnwidth, clip=true]{BPCIntroFig_ver6_sub3}}\quad\quad
  \subfigure[]{\label{fig:packetcountIntroduction_sub3}\includegraphics[viewport=0.1in 0.1in 5in 4in,width=0.6\columnwidth, clip=true]{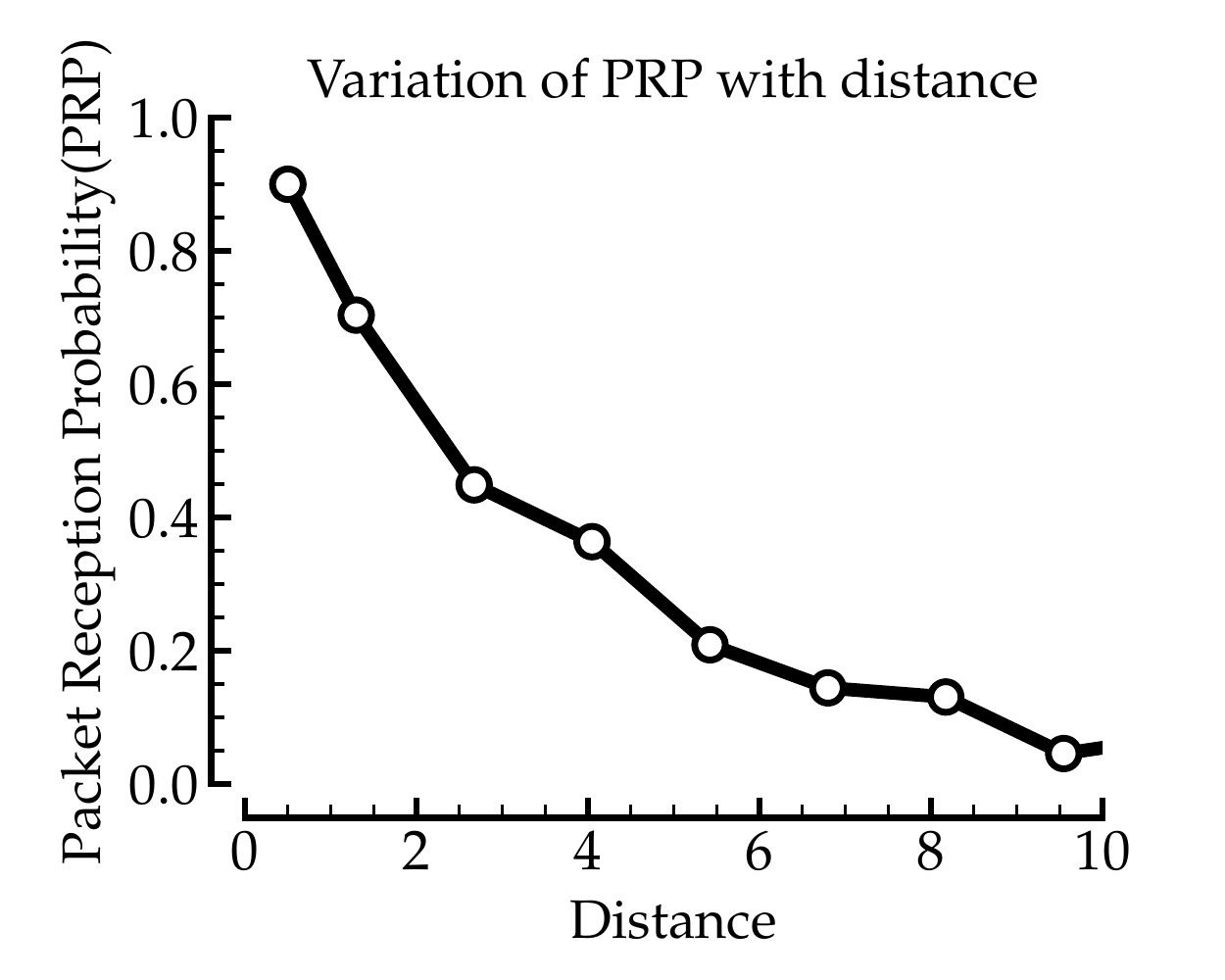}}
  \subfigure[]{\label{fig:packetcountIntroduction_sub4}\includegraphics[viewport=0.2in 0.2in 5.7in 6.2in,width=0.6\columnwidth, clip=true]{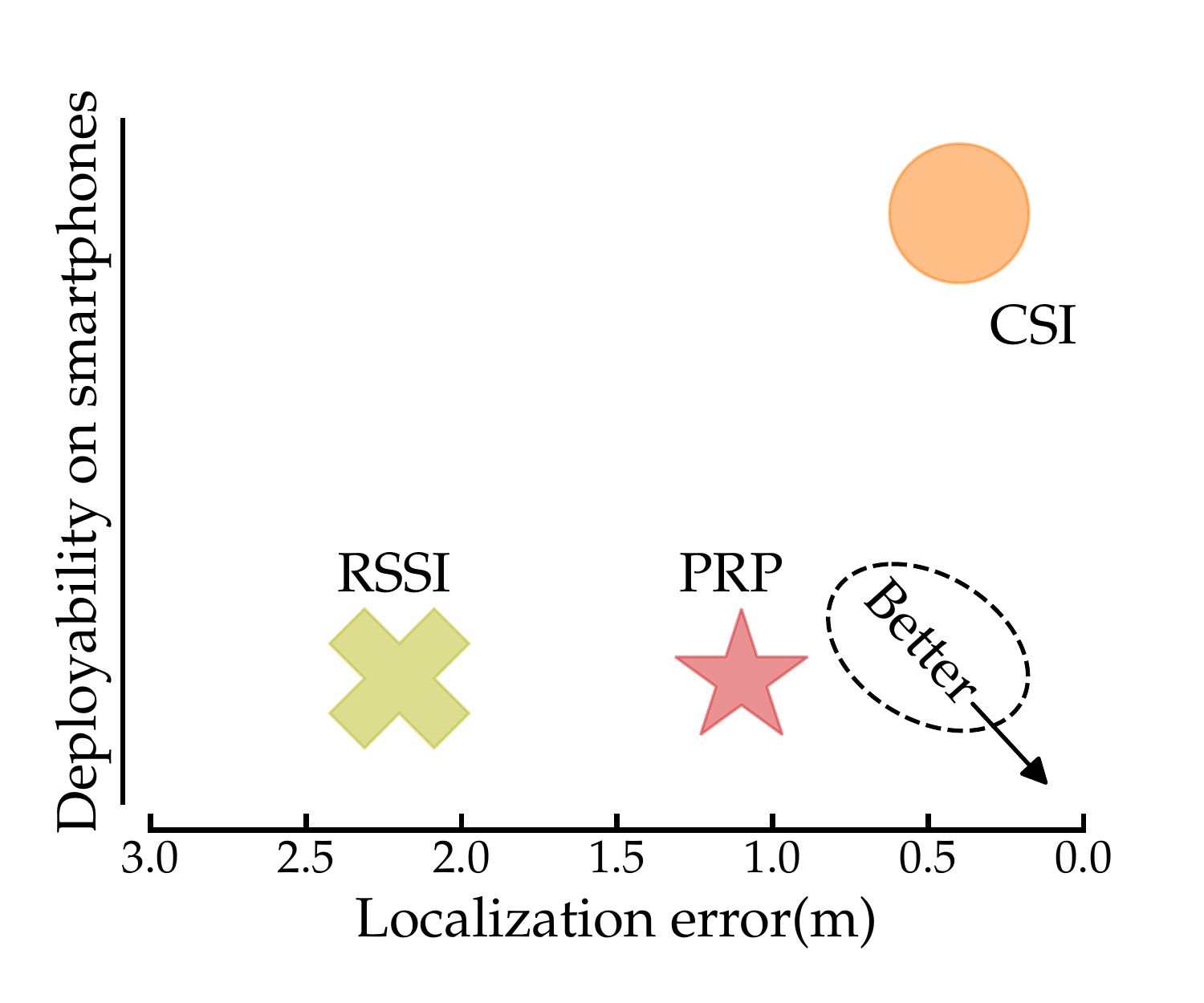}}
  \vspace{-0.1in}
  \caption{(a) As the mean RSSI decreases, the error in the RSSI estimate increases because of lost packets. (b) Packet reception in Line-of-Sight (LOS) with -20db transmission power decreases with distance. (c) Packet Reception Probability(PRP) technique is more accurate than RSSI~\cite{bahl2000radar, heurtefeux2012rssi, zanella2016best} and readily deployable on commercial devices than CSI~\cite{kotaru2015spotfi, xiong2013arraytrack, vasisht2016decimeter}.
  \vspace{-0.2in}
    % Multipath effects causes errors in distance estimates. (c)
  }
  % \vspace{-0.2in}
  \label{fig:packetcountIntroduction}
\end{figure*}

In this paper, we ask: \textit{Can we develop robust Bluetooth based contact tracing, with existing measurements, deployed on low-cost commodity hardware?} To do so, we need to overcome four fundamental limitations---deployability, bias in RSSI, high packet loss in Bluetooth and multipath effect.

\vspace{4pt}\noindent\textbf{Deployability on commercial smartphones:} Bluetooth Low-Energy based apps for contact-tracing have two well-known shortcomings. These apps primarily use either RSSI (Received Signal Strength Indicator) or presence to determine risk to COVID exposure . Prior work~\cite{bahl2000radar, heurtefeux2012rssi, zanella2016best} demonstrates that RSSI-based methods experience large errors (order of several meters) in positioning, especially in the low RSSI-large distance regime. RSSI has an important benefit: it is present on all modern devices.  In contrast, we cannot use CSI (Channel State Information)~\cite{kotaru2015spotfi, xiong2013arraytrack, vasisht2016decimeter}, a recent method that enables sub-meter accuracy, since off-the-shelf devices typically do not report CSI. A recent work ~\cite{schulz2018shadow} has enabled CSI for WiFi in some smartphones, but cannot be applied for BLE. Some contact-tracing apps also use `presence'---if one device can hear another---to determine if an individuals is close to another infected person. Presence is a poor proxy for distance since devices can hear Bluetooth beacons well beyond 6 ft social distancing radius, and also hear them across aisles and walls.
\vspace{4pt}\noindent\textbf{Biased RSSI Estimates due to Packet Loss:} We explain with a conceptual example in~\Cref{fig:packetcountIntroduction_sub1} that shows a Normally distributed RSSI at the receiver, for a fixed transmitter and receiver. In free space, with increasing distance between the transmitter and the receiver, the RSSI distribution shifts to the left, implying a decreasing RSSI at the receiver.
% As expected, the RSSI is Normally distributed with the variance caused by noise.
RSSI-based methods~\cite{bahl2000radar, madigan2005bayesian} empirically measure RSSI and use the mean RSSI estimate to infer distance. However, as the distance between the transmitter and the receiver increases (i.e., the RSSI distribution shifts to the left), packet loss increases with almost certain packet loss at the low-RSSI decoding threshold. Since devices only report RSSI for successfully decoded packets, RSSI-based distance methods suffer from a sampling bias: they use RSSI from decoded packets only. Since they cannot know RSSI values of packets they cannot decode, these methods introduce systematic error in their mean RSSI estimates. This error increases with distance, so much so that at large distances (few meters for BLE), as we shall show in this paper, the mean RSSI estimate becomes de-correlated with distance and is an unreliable indicator. The error is different from the typical reduction in SNR due to increase in distance. The error stems from a sampling bias fundamental to RSSI measurements.

\noindent\textbf{Packet Losses are higher in BLE:} Packet loss is a fundamental problem in a low-power protocol like Bluetooth Low Energy. At distances as small as \SI[]{1}{\meter}, in line of sight, around 10\% of the packets get dropped in our empirical evaluation, as shown in~\Cref{fig:packetcountIntroduction_sub3}. Packet loss rate increases to 50\% at \SI[]{3}{\meter}.
% Packet loss compromises the accuracy of RSSI measurements due to the sampling bias, and hence, the accuracy of distance estimates.
Thus, the sampling bias in RSSI measurements is a more  significant challenge for BLE compared to high-power WiFi protocol based RSSI methods~\cite{bahl2000radar, youssef2005horus, chintalapudi2010indoor}. As pointed out in ~\cite{chen2017locating}, BLE limits transmission power to reduce energy consumption. BLE v4.0, v4.1, and v4.2 defined maximum output power is 10mW, which is $10\times$ lower than WiFi.
% Though BLE v5.0 sets the maximum output power to 100mW, but the high Tx power is designed exclusively for high power devices with Class 1 BLE chip, and not for BLE beacons.

\noindent\textbf{Multipath Effects:} Multi-path effects~\cite{wen2015fundamental, zanella2016best} are the second large contributor to RSSI errors. Specifically, the error arises due to reflections of the radio signals by objects in the environment. Thus, the signals from the transmitter travel along multiple paths and combine at the receiver. This combination can be constructive (i.e., in-phase) and increase RSSI or destructive (i.e., out of phase) and reduce RSSI. Since this combination is a function of the environment and \textit{not} the distance between the devices, multipath introduces error in distance measurements.
% The multipath effect is well-documented~\cite{wen2015fundamental, zanella2016best} and is not a novel contribution of our work.
%\end{description}

% because they only average the RSSI for successfully received packets. As the distance becomes larger, this error keeps growing, so much so that at large distances (few meters in Bluetooth Low Energy), the RSSI becomes decorelated with distance.

% \vspace{-0.2in}
\subsection{A Counter-Intuitive Approach: Exploit Packet Loss to infer Distance}
\label{sub:Our Approach}

In this paper, we ask a counter-intuitive question: \textit{Could the loss of a packet be a clue to the distance between the transmitter and receiver?}
Intuitively, as the distance between transmitter and receiver increases, the ability to successfully receive packet decreases. In this paper, we build on this intuition to develop a new metric: Packet Reception Probability (PRP), which measures the probability that a receiver successfully receives packets from the transmitter.
% To see why $\mathrm{PRP}$ depends on distance, consider the packet loss rate $Q$, where $\mathrm{PRP} = 1 - Q$. Packet loss $Q$ depends on the RSSI distribution that depends on distance $x$ between transmitter and receiver and transmitter power $z$. Thus the expected packet loss, $\mathbb{E} [\mathrm{Q}_{x; z}] = \int_{-\infty}^{\infty}  P(Q \mid r_{x;z}) P(r_{x;z})\mathrm{d} r_{x;z}$, where $r_{x;z}$ is the RSSI random variable at distance $x$ and power $z$.
% PRP captures the information about lost packets.
% The fundamental problem here is that RSSI-measurements only capture information from packets that are received and fail to capture any information about the packets that have been lost. \textit{We ask if it is possible to incorporate information about lost pockets in the location estimation methods.}
% Specifically, we propose a new metric: Packet Reception Probability (PRP). PRP measures the probability that a receiver successfully receives packets transmitted by the transmitter. PRP captures the information about lost packets. In fact, we find that for low energy protocols like BLE, PRP is a good indicator of the distance between the communicating devices.
A simple experiment validates our intuition that PRP can encode distance. We collected packets from BLE beacons transmitting at -20db power at increasing distance values between \SIrange[]{1}{10}{\meter} in a line-of-sight (LOS) scenario. We use maximum likelihood estimates for PRP. We plot the PRP estimate as a function of distance in~\Cref{fig:packetcountIntroduction_sub3}. Notice that~\Cref{fig:packetcountIntroduction_sub3} shows that the probability of receiving a packet \textit{decreases} with distance, implying that PRP encodes distance. We show in this paper, that for low energy protocols including BLE, PRP is a good indicator of the distance between communicating devices.

Our approach, Bayesian Packet Reception Probability (B-PRP), is suitable for public spaces including retail stores or libraries, places that are important to current social distancing and contact tracing efforts. B-PRP is a PRP-based approach that develops a novel Bayesian framework to explicitly model multipath reflections in the environment and deliver robust and accurate localization. The Bayesian framework helps to minimize system deployment costs.
% Due to the simplicity of the packet reception framework, B-PRP can be deployed as an application on off-the-shelf commercial smartphones, unlike CSI-measurements which require firmware or hardware changes. While we evaluate PRP as a sole indicator of distance to highlight it's benefits, we show that B-PRP, when combined with RSSI, improves the performance of the system even further.
A public environment like a retail store contains obstructing materials in the form of stacks or shelves. The shelves (including the items placed on them) absorb or reflect the radio signals directed at them. This leads to a lower packet reception probability at the receiver. At a fixed distance, the packet reception probability will vary based on the number and type of obstacles in the signal path. B-PRP must tease apart the effects of distance from the interference effects of the obstacles, when estimating distance. We observe that we can model public spaces including retail stores in a modular manner comprising open spaces separated by stacks. We explicitly capture the effect of such stacks by modeling the packet reception in absence of stacks and in presence of one stack, two stacks and so on. While we use stacks to model retail spaces, we believe that the abstraction of modeling a geometric element is general enough to apply to other large indoor spaces like libraries, warehouses, factories, etc.

Finally, we present a method to estimate inter-device distance using our approach, a primitive essential to contact-tracing. We evaluated B-PRP in two real-world public places, an academic library setting and in a real-world retail store, and demonstrate the efficacy of our techniques.
% This makes B-PRP easy to deploy and use. We conducted extensive experiments in two test-beds: an academic library setting and in a real-world retail store.
In both cases, we did not control for human traffic. Our main results:

\begin{description}
  \item[Localization Accuracy:] B-PRP achieves a median localization error of \SI[]{1.03}{\meter} (library) and \SI[]{1.45}{\meter} (retail store). The state of the art Bayesian RSSI system ~\cite{madigan2005bayesian} has errors of \SI[]{1.30}{\meter} (library, 26.2\% more error) and \SI[]{2.05}{\meter} (retail store, 41.3\% more error) when trained with the same number of data points and packets per data point.
  \item[Distance estimation for contact tracing:] Our contact tracing distance estimation achieves median error of \SI[]{0.97}{\meter} (library) and \SI[]{1.22}{\meter} (retail store) with PRP values. The errors with RSSI are \SI[]{1.69}{\meter} (library, 74.2\% more error) and \SI[]{1.25}{\meter}(retail store, 2.4\% more error). Using the covid risk metric \cite{van2020aerosol}, we see that PRP does 1000X better than RSSI in the library.
  % \item[Peer to peer distance measurement:] B-PRP achieves distance estimates between two devices to an accuracy of \textbf{XX m}. In contrast, peer to peer distance measurements systems (like the one used in Apple-Google ENS) can achieve an accuracy of \textbf{YY m} in the same space.
  \item[B-PRP+RSSI Fusion:] Fusion of B-PRP and RSSI modestly improves the overall localization accuracy over B-PRP (\Cref{tab:PRPvsRSSItable}). We see best fusion results at small distances (\SI{\le2}{\meter}). At larger distances (\SI[]{\ge 2}{\meter}), errors in RSSI cause fusion results to be significantly worse than B-PRP. PRP+RSSI also improves contact tracing accuracy by 6\% for both library and retail store.
  \item[Robustness to Multipath:]   Our multipath model increases the accuracy for PRP from \SIrange[]{1.41}{1.03}{\meter} in the library (a 26.9\% improvement) and from \SIrange[]{1.60}{1.45}{\meter} (a 9.3\% improvement) in the retail store.
  \item[Number of Beacons:] As beacon density decreases, B-PRP error is always within $2m$ while RSSI errors are higher than $3m$. With five beacons, B-PRP performs $65\%$ better in library and $50\%$ better in the retail store.
  \item[Low Training Overhead:] B-PRP can leverage unknown training data to train the B-PRP model, thereby reducing the deployment effort. Specifically, B-PRP can achieve 1.08 m median accuracy with just 8 labelled data points and 4 unlabelled data points.
  %\item[Reduced Retraining Effort:] When trained with 8 known training spots, B-PRP achieves a median error of $1.82m$. The error reduces to $1.08m$ when 4 unknown training spots (no location information) are added. Thus, re-training B-PRP requires minimal effort.
\end{description}

For completeness, we note that the core limitation of a localization method like B-PRP, a limitation shared with methods including \cite{chen2012fm,lau2009measurement,grosswindhager2018salma,liu2017smartlight} is that it needs deployment of beacons in the public space to locate individuals.  However, BLE beacons are inexpensive, and our method, B-PRP, provides meter-level accuracy. A peer-to-peer distance estimation is much more general where we will use devices like smartphones for reception and transmission.
% but our experiments in~\hnote{Section XX} show that peer-to-peer RSSI-based distance measurements can go upto \textbf{XX} meters or more.
We believe that this tradeoff between some upfront infrastructure expense (multiple beacons) and increased localization accuracy is worthwhile in highly frequented public spaces.

\section{Contributions}\label{sec:contributions}
Our paper makes the following contributions:
\begin{description}
  \item[Use of Negative Information:]  To the best of our knowledge, we are the first to build an indoor positioning system that can extract information from \textit{absence of packets}. In contrast, state of the art RSSI based techniques~\cite{bahl2000radar, youssef2005horus, chintalapudi2010indoor}, use observed RSSI to infer distance. We accomplish this through a Bayesian formulation of the packet reception probability, a metric that we show encodes distance. We develop generic stacking models of reception to address multipath effects. While we use PRP as a sole indicator of distance to highlight its benefits, we show that B-PRP when combined with RSSI, improves the performance of the system at shorter distances. Our finding shows how to use BLE to robustly estimate indoor distances, thus opening the door to reliable BLE based contact-tracing that incorporates distance.
  \item[Distance estimation for contact tracing without localization: ] We directly estimate distance between two individuals \textit{without localization} by exploiting the well-known triangle inequality constraints in Euclidean geometry.
  % in our Bayesian framework, and construct a joint likelihood optimization problem that helps in better estimation of contact tracing distances.
  In contrast, we may consider estimating contact tracing distance through localization: that is, we first, estimate locations of two persons independently and then calculate the Euclidean distance between the two locations. This approach is sub-optimal---we are estimating location while we are only interested in distance. Also, if we have localization errors for a particular individual, these errors will impact \textit{all} the distance estimations between this individual and other nearby persons. We extend out Bayesian framework to independently estimate distances between pairs of individuals. With the known beacons, we form triangles and we impose triangle inequalities on these distances to rule out many distance configurations in the real world.  We improve our distance estimates by $\sim 10\%$ by triangle inequality distance estimation.
  \item[Sampling Bias in RSSI:] We show the effect of packet loss on the mean RSSI measurements. Furthermore, we show that with increasing distance, mean RSSI becomes highly unreliable due to sampling bias. Our finding is significant because the state of the art RSSI based techniques~\cite{bahl2000radar, youssef2005horus, chintalapudi2010indoor} when applied to BLE, a low-power protocol, are highly unreliable in the \SIrange[]{2}{6}{\meter} range (\Cref{tab:PRPvsRSSItable}). We highlight that \SI[]{2}{\meter} $\approx$ 6 ft, the social distancing range.
  \item[Readily Deployable Solution:] Our B-PRP framework does not require any hardware, firmware, or driver-level changes in off-the-shelf devices, and requires minimal deployment and re-training costs. In contrast, CSI~\cite{kotaru2015spotfi, xiong2013arraytrack, vasisht2016decimeter}, which can deliver sub-meter accuracy, requires firmware or hardware changes. This is significant: due to the simplicity of the packet reception framework, we can immediately deploy B-PRP as an application on off-the-shelf commodity smartphones.
\end{description}

\section{Motivation}
\label{sec:Motivation}

In this work, we focus on localizing individuals in indoor public spaces like retail stores, libraries. In these spaces, indoor positioning using BLE beacons can enable traditional applications like capturing behavioral data about shoppers, as well as novel applications like enforcing social distancing and contact-tracing. BLE offers an unique advantage for localization. Due to its low power budget, it can be turned on frequently and hence, enable more frequent location updates as compared to high power protocols like Wi-Fi. Recall that BLE's maximum transmit power (10 dBm) is 10 times lower than that of Wi-Fi (20 dBm). This factor, in addition with its ubiquitous presence on off-the-shelf smartphones, has made BLE the natural choice for such applications.

Traditional BLE localization techniques either use RSSI (Received Signal Strength Indicator) ~\cite{bahl2000radar, heurtefeux2012rssi, zanella2016best} or CSI (Channel State Information)~\cite{kotaru2015spotfi, BLoc}. CSI for BLE is not available on commercial devices like smartphones. On the other hand, RSSI measurements are noisy due to packet loss and multi-path effect. While multi-path effects~\cite{wen2015fundamental, zanella2016best} are well-documented, lets dig deeper into the challenge of packet loss.

First, we identity that packet losses can be mainly attributed to two reasons---random errors and low signal strength. Errors can occur uniformly at random irrespective of the RSSI of the packet. As a result, such errors do not introduce any bias in the aggregate estimate of RSSI. On the other hand, all packets that are received with a signal strength below a certain decoding threshold get dropped. Since we cannot observe the RSSI values of these low RSSI packets, and hence cannot include them in our aggregate estimates, we should expect to see a positive bias introduced in our RSSI measurements.

Lets mathematically validate our hypothesis of positive bias in RSSI aggregate estimates. Let us assume that the actual RSSI values at a certain location follow the Gaussian distribution $\mathcal{N} (\mu, \sigma^2)$. Lets further assume that the RSSI decoding threshold is $\alpha$. Since we drop all packets below the threshold, our aggregate RSSI estimates will be based on a Normal distribution truncated at $\alpha$. The new mean of this truncated normal distribution is given by
\begin{equation}
\hat{\mu} = \mu + \frac{\phi(\alpha)}{1 - \Phi(\alpha)} \sigma, \label{eq:truncated rssi mean}
\end{equation}
where, $\phi(\alpha)$ is the pdf of normal distribution evaluated at $\alpha, \phi(\alpha)\geq 0$. $\Phi(\alpha)$ is the cdf value of the normal distribution at $\alpha, \Phi(\alpha)<1$. Thus the estimate $\hat{\mu}$ that we obtain by measuring received RSSI values is biased by a positive amount of $\frac{\phi(\alpha)  \sigma }{1 - \Phi(\alpha)}$. As we move towards the lower RSSI regime, $\mu$ becomes closer to $\alpha$. As a result, both $\phi(\alpha)$ and $\Phi(\alpha)$ increases with lower RSSI values, which leads to a higher bias in the estimated mean RSSI.

Note that we cannot trivially estimate $\mu$ from $\hat{\mu}$ in ~\Cref{eq:truncated rssi mean} since in practice, multi-path effects alter the values of the RSSI in the received packets. Thus recovering $\mu, \sigma$ using say Maximum Likelihood Estimates by assuming a value of $\alpha$ is non-trivial.

Based on the above discussion, we identity that our solution requires three important properties---eliminating positive bias due to packet loss, robustness to multi-path effects, and ease of deployability on commercial devices. CSI meets the first two properties but misses the important requirement of deployability. RSSI is deployable, but has positive bias and is sensitive to multi-path.

% \input{packet-loss.tex}
% \input{problem-3}
%!TEX root = main.tex
\vspace{-0.1in}
\section{System Design}
\label{sec:Localization Problem}
In this paper, we solve the challenges with RSSI by asking a different question---\textit{Can we use the loss of packets as a signature itself to measure distance?} We define a random variable, packet reception probability, $prp(b)$, for a beacon $b$ whose expected value is defined as:
\begin{equation}\label{eq:prp}
    \mathbb{E}(prp(b))=\frac{\sum_i \bm{1}_{i=b}}{R(t_l-t_f)}
\end{equation}
Here, $\bm{1}$ is the indicator function that is $1$ if and only if packet $i$ is received from beacon $b$, $R$ is the sending rate of the beacon, and where $t_l$ and $t_f$ are the timestamps of the last and the first packet received from beacon $b$. Notice that the right hand side of~\Cref{eq:prp} is just the frequentist estimate of the probability of packet reception from beacon $b$: number of packets received divided by the total number of packets sent by beacon $b$.

One might wonder if PRP provides additional information beyond RSSI measurements. Notice that by directly modeling packet reception, we are leveraging absence of information(packet loss). RSSI is measured for packets that are successfully received, but not for dropped packets. Therefore, a system that drops 90\% of the packets and 50\% of the packets may have the same measured RSSI, but we know one of them has lower true RSSI, and hence is farther off, by looking at the packet reception probability. Also, \textbf{packets that are successfully received and influenced by multipath effects, only impact RSSI mean estimate but not expected PRP value $\mathbb{E}(prp(b))$}. Now we will focus on how to use PRP to measure location.

\subsection{Estimating Location using PRP}
\label{sec:Our Solution}

\begin{figure}[!htb]
  \centering
  \includegraphics[viewport=0.2in 0.2in 6.4in 4.0in, width=\columnwidth, clip=true]{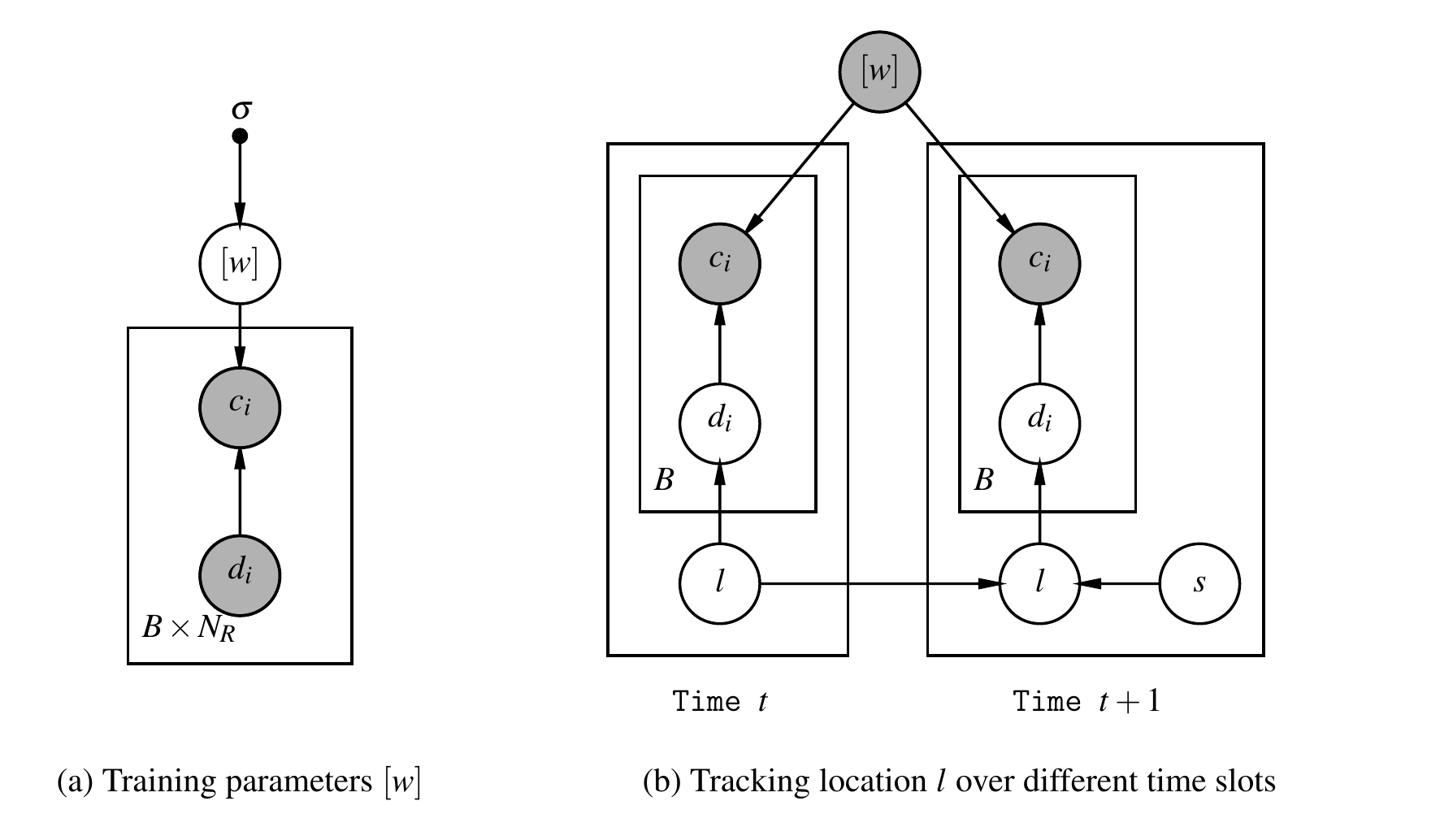}
  \vspace{-0.2in}
  \caption{ \textbf{Graphical model}: Shaded nodes are observed, while we need to estimate the unshaded ones. We use the data on number of received packets $c_{i}$ measured from $B$ beacons at $N_{R}$ reception locations to train the PRP parameters $[w]$.
%   We assume all beacon and reception positions to be known. We relax this assumption in ~\Cref{sec:Minimizing beacon set-up cost}.
  During tracking, we use the trained parameters $[w]$ and $c_{i,t}$ to estimate location $l_t$.
  % \vspace{-0.1in}
  }
  \vspace{-0.1in}
  \label{fig:plateNotation}
\end{figure}

\begin{figure*}[htbp]
    \centering
\begin{tikzpicture}[
roundnode/.style={circle, draw=green!60, fill=green!5, very thick, minimum size=2mm, scale=0.5},
squarednode/.style={rectangle, draw=red!60, fill=red!5, very thick, minimum size=2mm},
]

\fill[pattern=north east lines] (0,2.4) rectangle +(0.3,3.3); %walls
\fill[pattern=north west lines] (4.3,2.4) rectangle  +(0.3,3.3);

\fill[pattern=north east lines] (6,2.4) rectangle  +(0.3,3.3);
\fill[pattern=north west lines] (10.3,2.4) rectangle  +(0.3,3.3);

\fill[pattern=north east lines] (12,2.4) rectangle  +(0.3,3.3);
\fill[pattern=north west lines] (16.3,2.4) rectangle  +(0.3,3.3);

\draw (0,2.4) rectangle node[yshift=1.9cm] {\normalsize{Free Layout}} node[yshift=-2.0cm] {\normalsize{(A)}} +(4.6,3.3);  %boundaries
\draw (6,2.4) rectangle node[yshift=1.9cm] {\normalsize{Retail Layout}} node[yshift=-2.0cm] {\normalsize{(B)}} +(4.6,3.3);
\draw (12,2.4) rectangle node[yshift=1.9cm] {\normalsize{Geometric Elements}} node[yshift=-2.0cm] {\normalsize{(C)}} +(4.6,3.3);

\draw (7,5.2) rectangle node[xshift=-0.5cm] {\small Stack} +(2.5,0.3); %stacks
\draw (7,4.2) rectangle +(2.5,0.3);
\draw (7.5,3.2) rectangle +(2,0.3);

\draw (13,5.2) rectangle node[xshift=-0.5cm] {\small Stack} +(2.5,0.3); %stacks
\draw (13,4.2) rectangle +(2.5,0.3);
\draw (13.5,3.2) rectangle +(2,0.3);

\node[squarednode] (n1) at (3,5.35) [label=left:Beacon] {}; %beacons
\node[squarednode] (n2) at (9,5.35) {};
\node[squarednode] (n3) at (15,5.35){};

\node[roundnode] (n4) at (1.6,3.82) [label=below:Receiver] {1}; %receivers
\node[roundnode] (n5) at (7.4,4.82) {1}; %receivers
\node[roundnode] (n6) at (7.9,3.82) {2}; %receivers
\node[roundnode] (n7) at (9.0,2.82) {3}; %receivers
\node[roundnode] (n8) at (9.9,4.2) {4}; %receivers

\draw[->,line width=1.0pt,shorten >=0.1cm,shorten <=0.1cm] (n1) to node[align=center,xshift=0.9cm, yshift=-0.2cm] {Free\\ Space} (n4);
\draw[->,line width=1.0pt,shorten >=0.1cm,shorten <=0.1cm] (n2) to (n5);
\draw[line width=1.0pt,shorten <=0.1cm] (n2) to (8.4,4.5);
\draw[->,line width=1.0pt,shorten >=0.01cm,blue] (8.4,4.5) to (n6);
% \draw[->,line width=1.0pt,shorten >=0.01cm,shorten <=0.1cm,blue!70] (n2) to (n6);
\draw[line width=1.0pt,shorten <=0.1cm] (n2) to (9,4.5);
\draw[line width=1.0pt,blue] (9,4.5) to (9,3.5);
\draw[->,line width=1.0pt,shorten >=0.01cm,orange] (9,3.5) to (n7);
% \draw[->,line width=1.0pt,shorten >=0.01cm,shorten <=0.1cm] (n2) to (n7);
\draw[->,line width=1.0pt,shorten >=0.05cm,shorten <=0.1cm,brown] (n2) to (n8);

\fill[fill=white] (13,4.52) rectangle node {\small Free Space (F-S)} +(2.5,0.6);
\fill[fill=white] (13,3.52) rectangle node {\small 1-Stack Away(1-S)} +(2.5,0.6);
\fill[fill=white] (13.5,2.52) rectangle node {\small 2-Stack (2-S)} +(2,0.6);
\fill[fill=white] (12.33,3.22) rectangle node[rotate=90] {\small Corridor (C)} +(0.6,2.44);
\fill[fill=white] (15.54,2.42) rectangle node[rotate=90] {\small Corridor (C)} +(0.72,3.22);

\draw[dotted,line width=0.9pt] (13,4.5) to (13,5.2);
\draw[dotted,line width=0.9pt] (15.5,4.5) to (15.5,5.2);
\draw[dotted,line width=0.9pt] (13,3.5) to (13,4.2);
\draw[dotted,line width=0.9pt] (15.5,3.5) to (15.5,4.2);
\draw[dotted,line width=0.9pt] (13.5,2.4) to (13.5,3.2);
\draw[dotted,line width=0.9pt] (15.5,2.4) to (15.5,3.2);
\draw[dotted,line width=0.9pt] (12.3,3.5) to (13.5,3.5);

\draw[thin,double distance=2pt] (12.4,3.3) arc (90:0:0.8cm) node[midway,xshift=-0.3cm,yshift=-0.2cm,rotate=-45]{\small Desk}; %Desk
\draw (12.4,3.35) to (12.4,3.25);
\draw (13.15,2.5) to (13.25,2.5);

\draw[thin,double distance=2pt] (6.4,3.3) arc (90:0:0.8cm) node[midway,xshift=-0.3cm,yshift=-0.2cm,rotate=-45]{\small Desk}; %Desk
\draw (6.4,3.35) to (6.4,3.25);
\draw (7.15,2.5) to (7.25,2.5);

\end{tikzpicture}
\vspace{-0.1in}
\caption{\textbf{Modelling Obstacles and Multipath: } In (A), there is no obstruction in the path of the receiver. In retail layout (B), receiver 1 is in free space with beacon, 2 is one stack away and 3 is two stacks away. $4$ is an open region of the layout, i.e. the corridor.
We segregate the retail layout in (C) into geometric elements based on the relative position of beacon and receiver.}
\vspace{-0.2in}
\label{fig:learning}
\end{figure*}
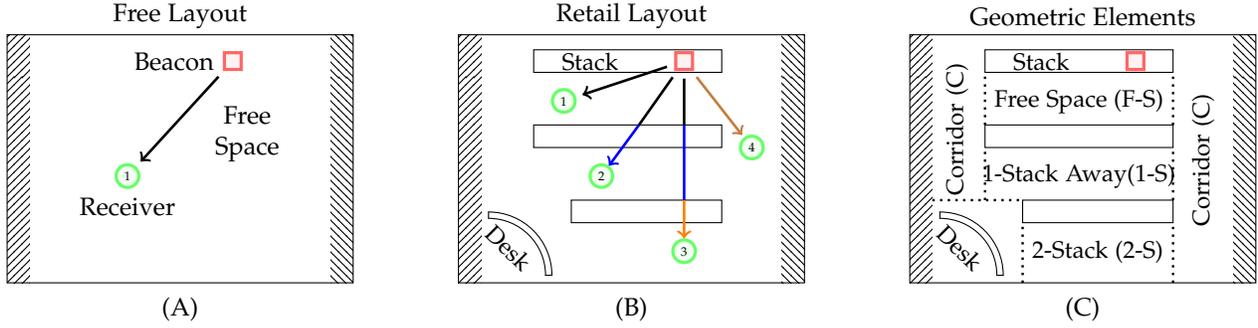

Recall, in ~\Cref{fig:packetcountIntroduction_sub3}, PRP degrades with distance. In this section, we discuss how we can model the relationship between PRP and distance, and use this relationship to infer location. Specifically, PRP ($prp$) depends on three factors: (a) distance ($d$), (b) sending rate ($R$), and (c) transmission power ($p_t$). In this subsection, we model the relationship in free space (~\Cref{fig:learning}(A)). We will incorporate the effect of multipath in subsequent sections.

We use a Bayesian model to model the relationship between PRP estimates from multiple beacons and the underlying physical location. Our choice of the Bayesian approach is motivated by two key design benefits: (a) It allows us to infer not just the location, but also quantifies the uncertainty in the location estimate. Such estimates are very helpful when the location is used for higher-layer applications like customer behavior analytics, contact-tracing.(b) It can be extended to  scenarios when the beacon location itself is unknown or the training set is small. As we show in ~\Cref{sec:Minimizing localization deployment overheads}, this reduces the deployment costs.

% Thus our data contains $B \times R \times \bar{T}$ packets, where $\bar{T}$ is the average time spent at each training location, received values, one for each pair of beacon and reception location.

% We view packet reception through the lens of a generative process.

We model $prp$ as a function $g$ of the distance $d$, sending rate $R$ and power $p_0$ of the beacon. Assume that we receive a packet from beacon $(x_b, y_b)$ at location $(x_r, y_r)$. We calculate the Euclidean distance $d$ between the beacon and receiver. Then, assuming that we know sending rate $R$ and transmission power $p_0$, we can model the number of packets received $c$ received at $(x_r, y_r)$ as drawn from a binomial distribution with parameter $prp$:

% The probability of packet reception $p$ is a function $g$ of the distance $d$ and frequency $f$ and power $p_0$ of the beacon.

\vspace{-0.25in}
\begin{align*}
    c &\sim Bin \left (N, prp\right),  & \textrm{binomial distribution}, \\
    prp &= g(d, R, p_0), & \textrm{PRP link function},\\
    d &= \sqrt{(x_b - x_r)^2 + (y_b - y_r)^2}, & \textrm{distance to beacon } b.
\end{align*}

% \vspace{-0.05in}
$N$ is the total number of packets sent out by the beacon is proportional to the product of the sending rate $R$, and the time spent $T_r$ at location $r$. The function $g(d,R,p_0)$ is a link function that connects the underlying infrastructure parameters ($R, p_0$) and physical distance $d$, to the packet reception probability.
% and depends detereministically on the frequency of the beacon.
% $N \propto f \times \bar{T}$ where constant of proportionality depends on the amount of time we hear to the beacon.

In identifying the right representation of $g$, we need to keep two considerations in mind: (a) the value of $g$ has to be between 0 and 1, and (b) $g$ must encapsulate relationship between $d$, $R$ and $p_0$, not just their direct effect on $prp$. Therefore, we model $g(d,R,p_0)$ as a logistic function of quadratic interaction between the parameters.

% More formally:
% \vspace{-0.2in}
\begin{equation}
    \text{logit}\{g(d,R,p_0)\} = w_0 + \sum_i w_i \theta_i + \sum_{i, j} w_{i,j} \theta_i  \theta_j \label{eq:linear model}
    % \log{\frac{g(d,f,r)}{1 - g(d,f,r)}}= b_0 + \sum_i b_i x_i + \sum_{i, j} b_{i,j} x_i  x_j \label{eq:linear model}
\end{equation}

\vspace{-0.05in}
where, $\text{logit}(p) = \log(p/1-p) $. And where, $\theta_1,\theta_2,\theta_3$ correspond to the variables of $d, R, p_0$ respectively. The coefficients $[w] = [w_i, w_{i,j}]$ are drawn from a non-informative prior $N(0,\sigma)$---a zero mean Normal distribution with variance $\sigma$. We choose $\sigma$ to be large in our system to allow for a large range of values.
% $\sigma(a) = 1/(1 + exp(-a)) $
% Our observed data $D$ in the model is the Packet Reception Probability $c$.

Our Bayesian formulation above is shown in ~\Cref{fig:plateNotation}(a). Our framework operates as follows:

\noindent\textbf{Training Phase: } During training, we use a data set $D$ collected in an environment to estimate the underlying parameters. Specifically, we need to estimate the posterior distribution of the unknown parameters $[w]$ given data $D$ i.e. $P([w] \mid D)$. The training set, $D$, comprises BLE logs. Specifically, to obtain $D$, we stand at $N_R$ locations in our testing area and listen to the packets from $B$ beacons. Assume further, that we know the $B$ beacon locations $(x_b, y_b)$, $b \in \{ 1,\dots,B\}$ and $N_R$ reception locations  $(x_r, y_r)$, $r \in \{ 1,\dots,N_R \}$. We will relax this assumption in ~\Cref{sec:Minimizing localization deployment overheads}.

\noindent\textbf{Test Phase: } During test phase, we do not know the reception locations, $(x_r, y_r)$ $r \in \{ 1,\dots,N_R \}$. We use the measured $prp$ and the parameters estimated during the training phase to estimate the receiver location. We use PyMC3 \cite{salvatier2016probabilistic} framework to do the inference.

\noindent\textbf{Adding Human Mobility: } Finally, we note that human location across time is not independent. Rather, locations are constrained by the time between them and the average moving speed of a person. If we wish to track individuals at temporal resolution $\delta$, and if a person reaches a location at time $t$ with speed $s_t$, we can constrain that location in terms of previous location at $t-1$
\begin{align*}
    s_t & \sim U(0, S_{max}), & \textrm{speed}, \\
    x_t\mid x_{t-1} &\sim \mathcal{N}(0, s_{t}*\delta), & x_t \textrm{ constrained by } s_t \times \delta, \\
    y_t\mid y_{t-1} &\sim \mathcal{N}(0, s_{t}*\delta), &  y_t \textrm{ constrained by } s_t \times \delta.
\end{align*}
Where, $S_{max}$ is a constant in our model denoting maximum movement speed of a human (similar to~\cite{de2017finding}). We estimate speed and location of a person from $prp$ data.
%With $f$ the beacon transmission frequency, and $\bar{T}$ as the average time spent at each location, we can decode at most $B \times R \times f \times \bar{T}$ packets.

\subsection{Combating Multipath Effect}
We have assumed a free-space propagation model so far, but real-world environments have obstacles. We observe that the main contributor to multipath effect in public spaces like retail stores (or libraries) are the stacks used to list products (or books) and to separate aisles. In such scenarios, the $prp$ value depends not just on the distance, but also on the number of stacks the signal has to cross. Crossing one stack is easier than crossing two and will cause fewer packet drops.
%These obstacles are mainly in the form of stacks that segregate the layout into a collection of geometric elements. For each geometric element, the obstruction scenario---the number of interfering stacks, surrounding walls---differs for each beacon and depends on the relationship between that beacon and the receiver.

To build on this observation, we explicitly model the number of stacks in our framework. This allows us to not just estimate the distance between a beacon and a receiver, but also estimate the number of stacks between them. Estimating this geometric information is useful for both: combating multipath, and exploiting in higher-layer applications. For instance, retail store apps need to estimate what aisle a customer is shopping in, contact-tracing apps want to discount for infection spread if customers are close (but across aisles). To estimate the stack separation, we divide the store layout in ~\Cref{fig:learning} into five portions based on the given beacon---free space(F-S), one stack (1-S), two stacks away (2-S), corridor (C) and desk (D). In the figure, the packets to receiver $1$ in F-S do not have to go through any obstacles. The packets to receiver $2$ in 1-S and $3$ in 2-S go through one and two interfering stacks respectively. Receiver $4$ is in a corridor. We limit ourselves to two stacks away in the model, because we empirically observe that two stacks or more have similar effects on packet reception (high loss).

Then, we parameterize our link function with a variable, $\gamma$ that denotes the geometric-element separation. We represent the new link function as $g_{\gamma}(d,R,p_0)$.
% We learn a separate packet reception function $g$ for each geometric element of the store. I
Now, at training time, we estimate parameters for the functions---free space $g_{F-S}$, one stack $g_{1-S}$, two stack $g_{2-S}$ and corridor (C) model $g_{C}$. We use a Bayesian training procedure similar to the free-space scenario. We segment our training data into the different scenarios, and use the segment-specific data to learn the parameters in each $g_\gamma$. For example, the data with one stack separation is used to train $g_{1-S}$. During testing, B-PRP uses the maximum likelihood model to identify the underlying location as well as stack separation.

At first blush, it might seem very complex to identify $\gamma$ for each of the $B$ beacons. We exploit the knowledge of the store geometry and beacon arrangements within the store to significantly reduce the number of unknowns. Assume that beacons $a$ and $b$ are in the same aisle adjacent to each other. Then, \textit{regardless of where the individual is, beacons $a$ and $b$ must have the same model type $\gamma$ with respect to the receiver.} Similarly, if beacons $a$ and $b$ are in neighboring aisles, and the model type $\gamma$ is $F-S$ for $a$, then $\gamma$ must be $1-S$ for $b$. Thus, given a location $x_t, y_t$, knowledge of store geometry and beacon arrangements help fix the model type for \textit{all} beacons, given the model type for \textit{any one} beacon.

\subsection{Estimating Distance for Contact Tracing}
\label{sub:Estimating Contact Tracing Distance}

%Now, we illustrate how we can measure the contact tracing distance between two persons using the PRP observations that each person's phone(i.e. the receiver) makes independently with the beacons installed in the environment. In our current framework, receivers only passively receive packets sent out by the beacons. The receivers do not broadcast packets, and hence there is no direct PRP observation on the distance between two receivers. However note that our framework can be trivially extended to the scenario where receivers actively broadcast packets into the medium, and combine those inter-receiver PRP values with receiver-beacon PRP values to get a better estimate of the contact tracing distance between two receivers.

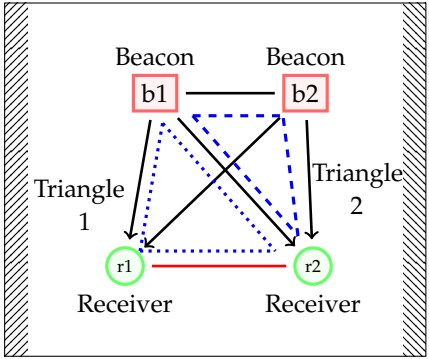
\begin{figure}[!htb]
\centering
\begin{tikzpicture}[
roundnode/.style={circle, draw=green!60, fill=green!5, very thick, minimum size=2mm, scale=0.7},
squarednode/.style={rectangle, draw=red!60, fill=red!5, very thick, minimum size=3mm},
]
\fill[pattern=north east lines] (0,1) rectangle +(0.3,4.7); %walls
\fill[pattern=north west lines] (5.3,1) rectangle  +(0.3,4.7);
\draw (0,1) rectangle +(5.6,4.7);
\node[squarednode] (n1) at (2,4.5) [label=above:Beacon] {b1}; %beacons
\node[squarednode] (n2) at (4,4.5) [label=above:Beacon] {b2}; %beacons
\node[roundnode] (n3) at (1.6,2.2) [label=below:Receiver] {r1};
\node[roundnode] (n4) at (4.1,2.2) [label=below:Receiver] {r2};
\draw[->,line width=1.0pt,shorten >=0.1cm,shorten <=0.1cm] (n1) to (n3);
\draw[->,line width=1.0pt,shorten >=0.1cm,shorten <=0.1cm] (n1) to (n4);
\draw[->,line width=1.0pt,shorten >=0.1cm,shorten <=0.1cm] (n2) to (n3);
\draw[->,line width=1.0pt,shorten >=0.1cm,shorten <=0.1cm] (n2) to (n4);
\draw [line width=1.0pt,shorten >=0.1cm,shorten <=0.1cm] (n1) to (n2);
\draw [line width=1.0pt,shorten >=0.1cm,shorten <=0.1cm,color=red] (n3) to (n4);
\draw[dotted,line width=1.2pt, color=blue](2.1, 4.1) to (1.8, 2.4);
\draw[dotted,line width=1.2pt, color=blue](2.1, 4.1) to (3.6, 2.4);
\draw[dotted,line width=1.2pt, color=blue](1.8, 2.4) to (3.6, 2.4);
\node[text width=1.2cm] at (1.0,3.2) {Triangle};
\node[text width=1.2cm] at (1.6,2.8) {1};
\draw[dashed,line width=1.2pt, color=blue](2.5, 4.2) to (3.9, 2.6);
\draw[dashed,line width=1.2pt, color=blue](2.5, 4.2) to (3.7, 4.2);
\draw[dashed,line width=1.2pt, color=blue](3.9, 2.6) to (3.7, 4.2);
\node[text width=1.2cm] at (4.7,3.4) {Triangle};
\node[text width=1.2cm] at (5.2,3.0) {2};
\end{tikzpicture}
\caption{Modeling contact tracing distance optimizing joint likelihood of observed PRP values and triangle inequalities}
\label{fig:contacttracing}
\end{figure}

Given our framework, we could trivially estimate the distance between two individuals  using a two-step approach: first, estimate their locations independently and second, calculate the Euclidean distance between the two locations.  Then, we could use this distance to ascertain whether two individuals were in contact for contact-tracing. However, this approach is sub-optimal. It requires us to determine four unknowns --- $(x,y)$ for the two devices, while we are only concerned about the final distance estimate between the two devices. Also, if we make errors in location estimation for an individual, that impacts all the distance estimations of this individual with other neighboring persons. In other words, we get correlated errors for independent distances between different pairs of individuals. \textbf{Can we do better?}

At a high level, we can improve the distance estimation process using two insights. \textit{First}, we don't need to model individual locations if we just care about distance. Therefore, we explicitly incorporate the distance between two devices as part of our Bayesian model. This helps us reduce the number of unknowns in our framework, and also helps to model the distance between each pair of individuals as an independent unknown. \textit{Second}, we leverage the triangle inequality. The triangle inequality states that given a triangle, the sum of two edges has to be greater than or equal to the third edge. This helps us rule out many triangular distance configurations. We present a detailed formulation of these insights below.

Finally, one might wonder: why do we need all this complexity? Why don't we just use the direct transmission between two devices to estimate distance---device A transmits to device B, device B measures PRP, and we convert that to distance? This approach would create a posterior distribution for distance, but with a large variance because interference by other nearby persons increases the uncertainty in our posterior distance distribution. To shrink this variance, we need other distance measurements, either to known beacons or to many other peers.

In this paper we adopt an infrastructure-assisted approach described above to help compute distances between pairs of individuals. Given that public indoor spaces like retail stores or restaurants (or even businesses) are more likely to be crowded, an infrastructure-assisted approach is reasonable in these settings.

\noindent\textbf{Approach: } We explain our approach using a toy example(pictorially represented in ~\Cref{fig:contacttracing}) which contains two beacons $b1$ , $b2$ and two receivers $r1$ , $r2$. We are interested in finding the distance $d_{r1,r2}$. The latent variables are $(d_{b1,r1}, d_{b1,r2}, d_{b2,r1}, d_{b2,r2})$, on which we have prp data. $d_{b1,b2}$ is known.

First, we infer latent variables $(d_{b1,r1}, d_{b1,r2}, d_{b2,r1}, d_{b2,r2})$ by constructing a joint likelihood function with two components---observed prp values, and triangle inequalities. Taking the receiver $r1$ as an example, we have the triangle $(r1,b1,b2)$ which gives us three triangle inequalities that can be converted to likelihood values as
\begin{align}
  L_T = \log P(d_{r1,b1} + d_{r1,b2} - d_{b1,b2} > 0) \nonumber\\ + \log P(d_{r1,b1} + d_{b1,b2} - d_{r1,b2} > 0) \nonumber\\
  + \log P(d_{r1,b2} + d_{b1,b2} - d_{r1,b1} > 0) \label{eq:TriangleInequality}
\end{align}

For estimating the latent variables involving the receiver $r1$, we can write down the joint log likelihood function as:
\begin{align}
  \max [\log P(prp_{r1,b1} | d_{r1,b1}) + \log P(prp_{r1,b2} | d_{r1,b2}) + L_T]\nonumber
\end{align}
% \begin{align}
%   \max_{(d_{r1,b1}, d_{r1,b2})} [\log P(prp_{r1,b1} | d_{r1,b1}) + \log P(prp_{r1,b2} | d_{r1,b2}) + L_T]\nonumber
% \end{align}

Second, we infer $d_{r1,r2}$ by maximizing the likelihood of triangle inequalities involving triangles with two receivers and one beacon. We have two triangles $T_1 = (r1,r2,b1)$ and $T_2 = (r1,r2,b2)$. We can construct the likelihood functions for $T_1$ and $T_2$ similar to ~\Cref{eq:TriangleInequality}.
We maximize
% \begin{align*}
%     L_{T_1} = \log P(d_{r1,r2} + d_{r1,b1} - d_{r2,b1} > 0) \\ + \log P(d_{r1,r2} + d_{r2,b1} - d_{r1,b1} > 0) \\ +  \log P(d_{r1,b1} + d_{r2,b1} - d_{r1,r2} > 0).
% \end{align*}Similarly we can write for $T_2$.
\begin{align}
    L = \max_{(d_{r1,r2})} [ L_{T_1} + L_{T_2} | d_{b1,r1}, d_{b1,r2}, d_{b2,r1}, d_{b2,r2}].\nonumber
\end{align}
% \begin{align*}
%     L_{T_1} = \log P(d_{r1,r2} + d_{r1,b1} - d_{r2,b1} > 0) \\ + \log P(d_{r1,r2} + d_{r2,b1} - d_{r1,b1} > 0) \\ +  \log P(d_{r1,b1} + d_{r2,b1} - d_{r1,r2} > 0), \\
%     L_{T_2} = \log P(d_{r1,r2} + d_{r1,b2} - d_{r2,b2} > 0) \\ + \log P(d_{r1,r2} + d_{r2,b2} - d_{r1,b2} > 0) \\ + \log P(d_{r1,b2} + d_{r2,b2} - d_{r1,r2} > 0),\\
%     L = \max_{(d_{r1,r2})} [ L_{T_1} + L_{T_2} | d_{b1,r1}, d_{b1,r2}, d_{b2,r1}, d_{b2,r2}].
% \end{align*}

We use PyMC3 potentials to construct these joint likelihood functions and then apply MCMC sampling techniques to solve them. We can also use RSSI or PRP+RSSI instead of PRP in our likelihood functions, which will serve as our different methods in ~\Cref{sub:Contact Tracing Accuracy Evaluation}.

\section{System Deployment and Optimization}
\label{sec:Minimizing localization deployment overheads}
To summarize, the B-PRP system operates in following steps:
\begin{itemize}
\item \textbf{Deployment: }We deploy BLE beacons at known locations in an environment like a retail store. The location of the beacons as well as the floor plan is uploaded to a B-PRP server. The server can reside on the cloud or be an edge device local to each environment.
\item \textbf{Training: }A user walks to fixed locations in the store with a smartphone app or another BLE receiver and measures the PRP values. The PRP values are uploaded to a server. The server uses these labelled PRP values, the beacon locations, and the floor plan to train the B-PRP model.
\item \textbf{Localization and contact tracing: }Finally, when new users walk in, they measure PRP for beacons already deployed in the store. The app on the smartphone uploads the PRP values to the server. The server uses the trained model to infer location of the users and sends it back to the user. The server also uses the PRP values from multiple users to infer the proximity distance between them. Note that, this system is centered on the user. If the user chooses not to share the PRP values with the server, no location estimation and contact tracing can be performed. Furthermore, the design also conserves power on the smartphone because the user never has to transmit any BLE packets. 
\end{itemize}

Finally, we transmit beacons using BLE advertising mode. This prevents the need for making any explicit connection between the user device and the beacon. The user device can ignore the advertising beacons to avoid localization.

% \subsection{Reducing Deployment Overhead}\label{sec:Minimizing localization deployment overheads}

\vspace{4pt}\noindent\textbf{Reducing Deployment Overhead: } Deploying the localization infrastructure has two major overheads---setting up the beacons at exact locations, and training. Knowing the location for beacons deployed by a large store is labor intensive. Similarly, training involves standing at multiple known locations inside the layout and collecting data for certain period of time. We ask two questions---\textit{1. Instead of costly human labor, can we infer most beacon locations from training data?  2. Can we leverage data from unlabeled locations of store workers to train our model?}

As it turns out, we can affirmatively answer both these questions in our formulation. We can leverage unlabelled data (without location information) that is collected by store workers as they move around the store to help train the model as well as to infer most beacon locations. We use data collected by store workers $D$ to solve both problems. $D$ contains number of packets received from all $B$ beacons at all $N_R$ training locations. Let us assume that we know the locations of a small number $b \ll B$ \textit{primary} beacons, with the remaining $B-b$ beacon locations unknown; Ideally we will like $b$ to be as close to $0$ as possible. Also assume that only a small number $r \ll N_R$ locations are known, with the remaining $N_R-r$ locations unknown. Our goal is to infer $B-b$ beacon and $N_R-r$ training locations from $D$ along with the packet reception model parameters $[w]$.

To enable this, we view the model through a generative process. We initialize the $(B-b)$ beacon and $(N_R-r)$ unknown reception locations from a uniform prior over the testing area which is of dimension $W \times L$. We want to jointly estimate the distribution of the unknown beacon locations $\{l_j \}, {j\in \{1, \dots, B-b \}}$, $ \{l_k \}, {k\in \{1, \dots, N_R-r \}}$ and packet reception model parameters $[w]$, given data $D$. In other words, we want to estimate the posterior distribution $P([l_j,l_k, w] \mid D)$. This can be easily achieved, given the Bayesian nature of our model. We use standard \textbf{Markov Chain Monte Carlo (MCMC)} based Bayesian inference techniques to compute the posterior distribution over the unlabelled data points and beacons. We use \textbf{No-U-Turn sampling (NUTS)}~\cite{hoffman2014no} included with PyMC3~\cite{salvatier2016probabilistic} to perform MCMC sampling. Therefore, B-PRP can leverage unlabeled data as well as unlabelled beacon locations to improve its location estimates and reduce the deployment overhead.

%!TEX root = main.tex
% \vspace{-0.17in}
\section{Experimental Set Up}
\label{sec:Experimental Test-Bed}
We evaluate B-PRP in two testbeds---an academic library and a retail store. Both spaces have shelves segregating the floor space into rectangular regions, i.e. aisles and corridors. The two environments differ in three main aspects---difference in layout, i.e. arrangement of rectangular areas and the presence of walls around the space, difference in material of shelves, and human interference. The retail store had more dynamic customer traffic movement during the experiments.
%Human traffic increases the chance of not hearing a packet at the reception location. While we did not restrict movement, the library space had less traffic in comparison.

\begin{figure*}[htb]
  \centering
  \subfigure[Library] {\label{fig:picLibrary}\includegraphics[viewport=0.2in 2.1in 10.8in 6.4in,width=0.7\columnwidth, clip=true]{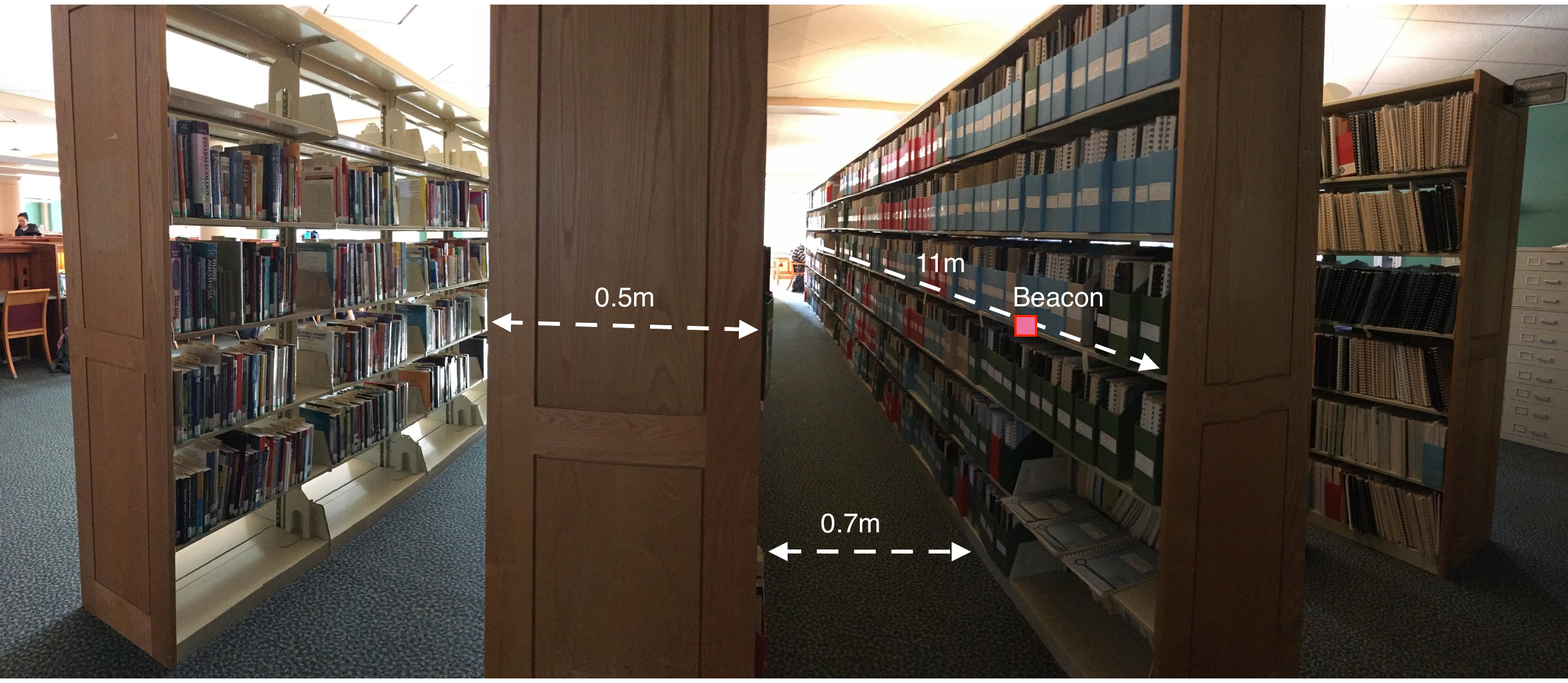}}\quad\quad
   \subfigure[Retail Store]{\label{fig:picRetail}\includegraphics[viewport=0.7in 1.6in 10.8in 6.5in,width=0.6\columnwidth, clip=true]{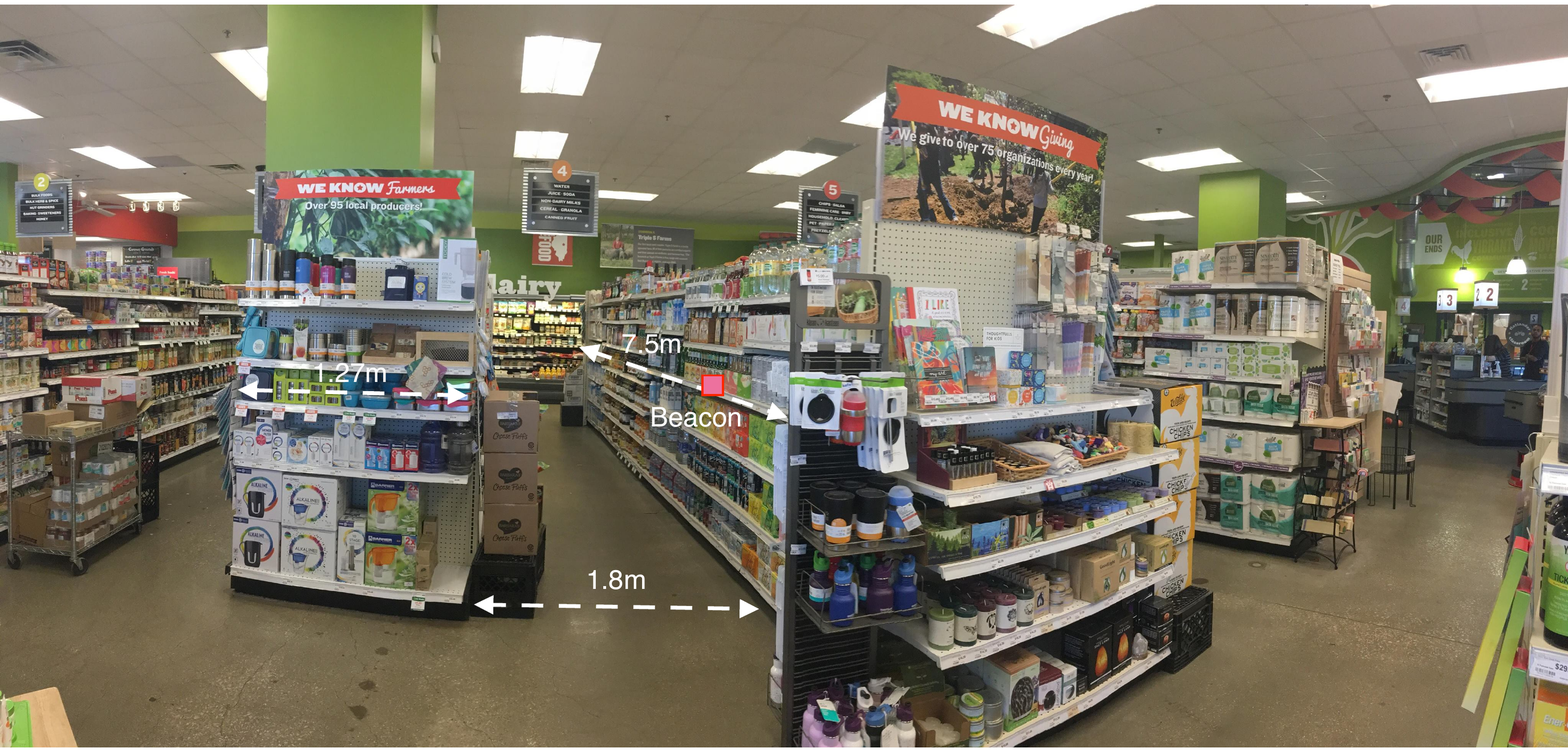}}\quad\quad
   \subfigure[Devices]{\label{fig:device}\includegraphics[viewport=0.6in 0.4in 12.1in 7.2in, width=0.5\columnwidth,clip=true]{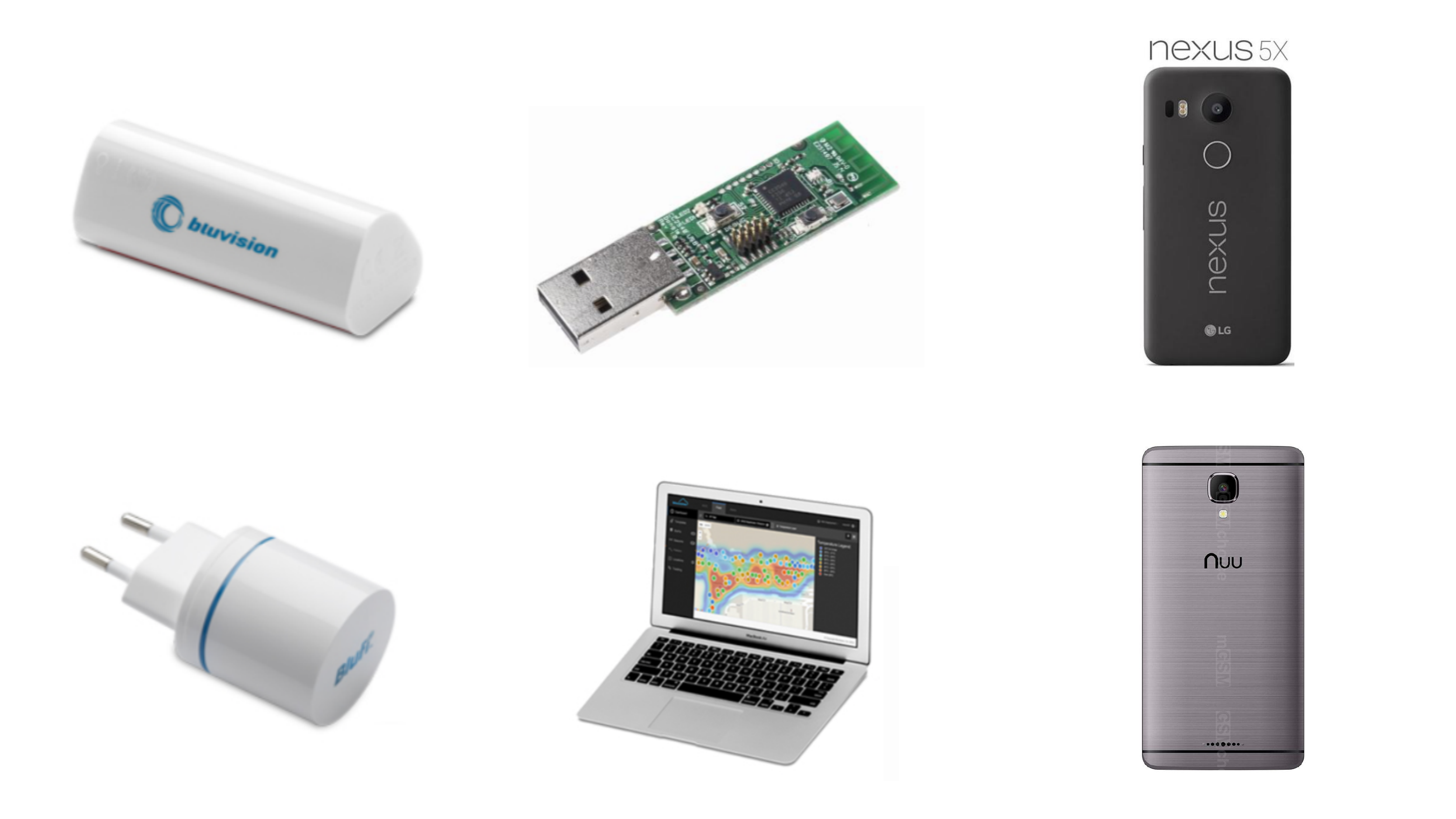}}
   % \subfigure[Devices]{\label{fig:device}\includegraphics[viewport=0.02in 3.1in 10.3in 5.3in, width=0.5\columnwidth,clip=true]{Devices2}}
  %\begin{subfigure}[t]{0.5\columnwidth}
  %   \centering
  %   \includegraphics[viewport=0.2in 2.1in 10.8in 6.4in, width=0.8\columnwidth,clip=true]{library2}
  %   \caption{Library.}
  %   \label{fig:picLibrary}
  %\end{subfigure}%
  %\begin{subfigure}[t]{0.5\columnwidth}
  %  \centering
  %  \includegraphics[viewport=0.7in 1.6in 10.8in 6.5in, width=0.8\columnwidth,clip=true]{retail2}
  %  \caption{Retail store.}
  %  \label{fig:picRetail}
  %\end{subfigure}
  \vspace{-0.15in}
  \caption{\textbf{Experimental Testbed: } We conduct our experiments in a library (a) and a retail store (b) using devices shown in (c)---Beacon, Blufi, Sniffer, Laptop, Nexus5X and NuuA4L android smartphone.}
  \vspace{-0.2in}
\end{figure*}

\textbf{Library:} We show the layout of the library space, $14m$ by $8m$, in
% ~\Cref{fig:imagebLocGrainger} and
~\Cref{fig:picLibrary}. It has three wooden shelves (each $11m$ long \& $0.5m$ wide). The aisles between two stacks are $0.7m$ wide. We placed two rows of 12 beacons on each stack. We manually measured each inter-beacon distance. The distance between two adjacent beacons on the same row is $0.91m$. The distance between two devices kept opposite each other on the same shelf, but facing two different aisles is $0.43m$. We carried out our experiments during regular library hours.

\textbf{Retail Store:}
~\Cref{fig:picRetail} shows a retail store with dimensions: $10m$ by $10m$. The environment has four steel stacks ($1.27m$ wide each; three are $7.5m$ long, one is $6m$ long). The aisles between two stacks are $1.8m$ wide. We place two rows of beacons on each stack. The inter-beacon distance on the same row is $1m$. Retail store is a challenging environment due to the presence of steel structures as well as worker and customer movement during the experiments. %We carried out most of the experiments in the grocery store during early morning hours. During regular hours, the customer traffic was high and the beacons were often displaced due to customers inadvertently moving them around. To ensure stable beacon locations and to avoid inconveniencing customers, we collected data in the morning. This was still a harsh environment: there were fewer customers, but we had store workers moving around arranging items in the aisle.

% \begin{figure}[htb]
%    \centering
%    \includegraphics[viewport=0.8in 1.3in 11.1in 11.3in, width=\columnwidth,clip=true]{beaconlocations_grainger}
%    \caption{Library layout. We show one specific movement sequence used in our experiments.}
%    \label{fig:imagebLocGrainger}
% \end{figure}

% \begin{figure*}[htb]
%    \centering
%    \includegraphics[viewport=1.7in 1.1in 12.1in 6.6in, width=\columnwidth,clip=true]{exptLayout}
%    \caption{Library and Retail Store layout. We show one specific movement sequence used in our experiments.}
%    \label{fig:imagebLocExpt}
% \end{figure*}

% \begin{figure}[t]
%    \centering
%    \includegraphics[viewport=0.7in 1.6in 10.8in 6.9in, width=\columnwidth,clip=true]{retail1}
%    \caption{Retail store Picture.}
%    \label{fig:picRetail}
% \end{figure}
%
% \begin{figure}[t]
%    \centering
%    \includegraphics[viewport=0.2in 2.1in 10.8in 6.6in, width=\columnwidth,clip=true]{library1}
%    \caption{Library Picture.}
%    \label{fig:picLibrary}
% \end{figure}

% \begin{figure}[htbp]
%     \centering
%     \includegraphics[viewport=1.2in 5.1in 7.1in 9.2in, width=\columnwidth,clip=true]{coop_layout}
%     \caption{Retail floor plan. We placed 38 beacons on the 4 stacks within the red square region.}
%     \label{fig:layoutCoop}
% \end{figure}
%
% \begin{figure}[htbp]
%     \centering
%     \includegraphics[viewport=1.8in 1.0in 9.1in 8.0in, width=\columnwidth,clip=true]{beaconlocations_coop}
%     \caption{Retail store layout. We show one movement sequence. Beacons are placed on stacks $1$ through $4$.}
%     \label{fig:imagebLocCoop}
% \end{figure}
%
% \begin{figure}[htbp]
%   \centering
%   \includegraphics[viewport=0.02in 3.1in 10.3in 5.3in, width=\columnwidth,clip=true]{Devices2}
%    \caption{Devices---Beacon, Blufi, Laptop and Sniffer}
%    \label{fig:device}
% \end{figure}

%

% \begin{figure}[htbp]
%     \centering
%     \includegraphics[viewport=0.8in 1.3in 11.1in 11.3in, width=\columnwidth,clip=true]{beaconlocations_coop}
%     \caption{Retail store layout. We show one specific movement sequence used in our experiments.}
%     \label{fig:imagebLocCoop}
% \end{figure}

% \begin{figure}[htbp]
%     \centering
%     \includegraphics[viewport=1.8in 1.0in 9.1in 8.0in, width=\columnwidth,clip=true]{beaconlocations_coop}
%     \caption{Retail store layout. We show one movement sequence. Beacons are placed on stacks $1$ through $4$.}
%     \label{fig:imagebLocCoop}
% \end{figure}
\vspace{-0.1in}
\subsection{Devices}
\label{sub:Devices}

% \begin{figure}[htb]
%     \centering
%     \includegraphics[viewport=1.2in 5.1in 7.1in 9.2in, width=\columnwidth,clip=true]{coop_layout}
%     \caption{Retail floor plan. We placed 38 beacons on the 4 stacks within the red square region.}
%     \label{fig:layoutCoop}
% \end{figure}

We use following devices for our experiments---Bluvision iBeeks~\cite{iBeek} , BluFi~\cite{bluFi}, TI packet sniffer, a laptop and Android smartphones(Nexus5x, NuuA4L). iBeeks or iBeacons are battery operated BLE beacons. They support a wide range of broadcasting power from $-40dBm$ to $+5dBm$. $-40dBm$ translates to $3m$ line of sight range, while $+5dBm$ gives us a range of $150m$. For our experiments, the beacons send 10 packets per second at -15 dBm power. We deploy 60 iBeacons in the library and 38 beacons in the retail store.

% \begin{figure}[htb]
%     \centering
%     \includegraphics[viewport=1.2in 5.1in 7.1in 9.2in, width=\columnwidth,clip=true]{coop_layout}
%     \caption{Retail floor plan. We placed 38 beacons on the 4 stacks within the squared region.}
%     \label{fig:layoutCoop}
% \end{figure}

% \begin{figure}[htbp]
%     \centering
%     \includegraphics[viewport=1.8in 1.0in 9.1in 8.0in, width=\columnwidth,clip=true]{beaconlocations_coop}
%     \caption{Retail store layout. We show one movement sequence. Beacons are placed on stacks $1$ through $4$.}
%     \label{fig:imagebLocCoop}
% \end{figure}

% \begin{figure}[htbp]
%   \centering
%   \includegraphics[viewport=1.6in 0.7in 11.2in 6.4in, width=\columnwidth,clip=true]{Devices3}
%    \caption{Devices---Beacon, Blufi, Laptop and Sniffer}
%    \label{fig:Test}
% \end{figure}

% \begin{wrapfigure}{l}{\columnwidth}
%   \centering
%   \includegraphics[viewport=0.02in 3.1in 10.3in 5.3in, width=0.5\columnwidth,clip=true]{Devices2}
%    \caption{Devices---Beacon, Blufi, Laptop and Sniffer}
%    \label{fig:device}
% \end{wrapfigure}

%\begin{figure}[htb]
%  \centering
%  \includegraphics[viewport=0.02in 3.1in 10.3in 5.3in, width=\columnwidth,clip=true]{Devices2}
%   \caption{Devices---Beacon, Blufi, Laptop and Sniffer.
%   \vspace{-0.1in}
%   }
%  \label{fig:device}
%\end{figure}

% \vspace{-15pt}
%BluFi enables mass re-configuration of iBeacons. It helps us to set the parameters, i.e. frequency and power of all the beacons. We can use Bluzone iOS app to re-configure beacons one at a time. BluFi can reconfigure a batch of beacons at once.

We use three receiver devices for BLE: Texas Instrument Packet Sniffer (CC2540 dongle), Nexus 5X smartphone, NuuA4L smartphone. iBeacons broadcast BLE packets in three channels--- 37, 38 and 39. The sniffer can filter out packets from specific channels. We connect the sniffer to a Windows laptop and use it for packet reception from beacons. For the Android phones, we built an android app using Altbeacon \cite{altbeacon} library to scan BLE channels.

\vspace{-0.1in}\subsection{Baselines}
\label{sub:Baselines}
We compare B-PRP against state-of-the-art in RSSI-based positioning:
\begin{itemize}
\item \textbf{Horus} \cite{youssef2005horus} is an RSSI fingerprinting technique that was originally tested with WiFi. We extend it to BLE. For fairness, we use Horus with the same number of training locations as other baselines---12 for library and 9 for retail store. The inter-state distance is $3.5m$ for library and $1.85m$ for retail store.
\item \textbf{Bayesian RSSI} \cite{madigan2005bayesian} uses a generative model based on RSSI to determine location. We set the priors and parameter values following recommendations in \cite{madigan2005bayesian}.
\item \textbf{Bayesian RSSI Fingerprinting} (or Bayesian FP) \cite{chen2013bayesian} is a Bayesian Fusion technique applied to a fingerprinting based method for BLE devices. It stores fingerprints like Horus, but employs fusion technique to combine current RSSI and prior location information.
\item\textbf{MCL} \cite{hu2004localization} is a range-free localization technique and uses proximity rather than ranging information to localize nodes. It observes whether a packet was received from a device and infers whether the reception location is inside or outside a threshold distance from the beacon.
\end{itemize}

To ensure fair comparison, we use the same training data across all techniques. Furthermore, for RSSI based techniques, we use mean RSSI values over all packets used by the PRP technique. That is, \textbf{if PRP uses $k$ packets at a location, we use the mean RSSI value over the \textit{same} $k$ packets.} This removes inter-packet RSSI variance at the same location, \textit{improving} RSSI localization. RSSI results are significantly worse without averaging.

There is more recent work in CSI-based positioning \cite{kotaru2015spotfi,xiong2013arraytrack,vasisht2018duet}, but CSI data is not available on most commercial smartphones. Hence, we do not cover these baselines. For reference, the state-of-the-art CSI-based method achieves a median localization error of 86 cm\cite{BLoc}. However, this work requires CSI data on phones and multi-antenna beacons, both of which are not mainstream yet, and hence, cannot be deployed at scale for applications like contact tracing.

\subsection{Data Collection}
\label{sub:Data Collection}

We collected data for both layouts in two phases---training and localization. We collected data at stationary spots to train B-PRP and competing baselines. We marked some fixed places for each layout and stood there for 1 minute to receive data from the beacons. We used 12 such spots for the library layout and 9 locations for the retail store layout.
%The number of locations depended on the area of the layout and the number of aisles. The data or packets collected from beacons include their MAC address and RSSI value. The MAC address uniquely identifies the beacon. We use the beacon identities for our framework and the RSSI information for the baselines.

% \begin{figure}[htb]
%     \centering
%     \includegraphics[viewport=0.9in 0.05in 13.1in 5.9in,width=\columnwidth, clip=true]{accuracyResults}
%     \caption{ CDF distribution for comparing tracking errors of Bayesian PRP for library and retail store. We can see that Bayesian PRP performs best in both cases. SAB-RSSI performs well in the library environment.}
%     \label{fig:imageResults}
% \end{figure}

 We collected data in both test-beds to compare the accuracy of localization and contact tracing techniques. To track and test on data from a moving person, we asked users to naturally move inside the layout with the laptop and sniffer in hand. We used fixed movement paths and marked spots along the path. Each path or trace is a simulated movement carried out in real time between such marked spots. We stop at each marked place for $10$ seconds, and we move at a normal walking speed of $0.5 m/sec$ between the spots. We can now calculate the ground truth location at any time within the movement trace. Please note that \textbf{we evaluate our location estimates throughout the movement trajectory}. They are not restricted to the marked fixed spots.
\vspace{-0.05in}
\section{Micro Benchmark}
\label{sec:MicroBenchmark}
\begin{figure*}[htb]
	\centering
	\includegraphics[viewport=3.2in 0.05in 19.0in 3.9in,width=2\columnwidth, clip=true]{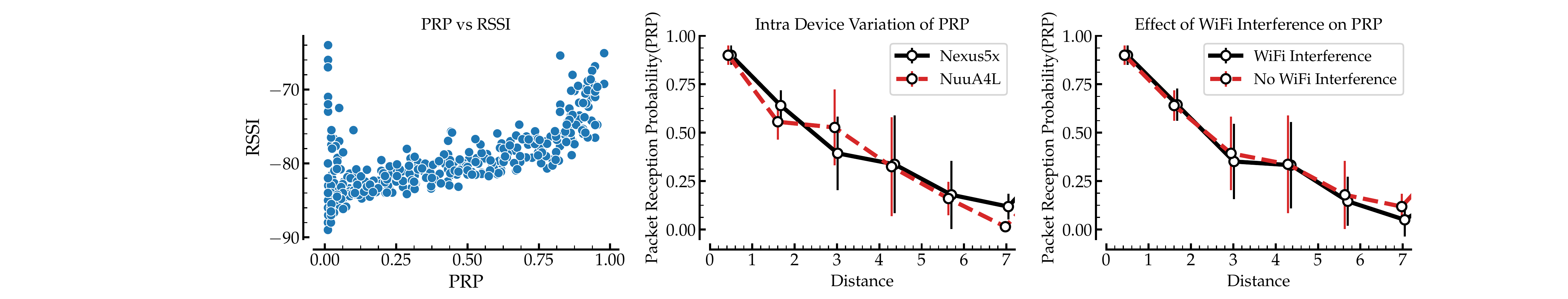}
  \vspace{-0.15in}
	\caption{ \textbf{Microbenchmarks: }(left) PRP vs RSSI are not directly related, but follow an expected trend. (middle) PRP variation is similar across two different android devices---Nexus5X and NuuA4L. (right) PRP is robust to ambient WiFi interference.}
	\vspace{-0.2in}
	\label{fig:benchMark}
\end{figure*}

We present microbenchmarks to better understand PRP:

\vspace{6pt}\noindent\textbf{Relationship to RSSI: }First we ask how PRP varies with RSSI and if packet reception is directly dependent on RSSI. We plot this relationship in ~\Cref{fig:benchMark}. As seen in the figure, there is an expected trend between the two parameters, but there is also significant variance for each value of PRP. This implies that the relationship between packet reception and RSSI is not determined by a hard threshold, but is instead more probabilistic. The probability of packet reception goes down with RSSI but several other factors including random noise come into play.

\vspace{6pt}\noindent\textbf{Translation across devices: }Does the relationship of PRP with distance depend on a device? To answer, we collect PRP values at the same location with two android smartphones: Nexus5X and NuuA4L. As shown in ~\Cref{fig:benchMark}, we see very close trends in PRP vs distance with minor variations\footnote{The experiments were conducted on different days for each smartphone}.

\vspace{6pt}\noindent\textbf{Robustness to interference: }Does interference from other in-band transmissions like WiFi hurt PRP? To understand this, we conduct the following experiment. We setup a WiFi router on 2.4GHz WiFi band and use two laptops to saturate the link using the iperf utility\cite{iperf}. We measure the PRP-distance relationship with WiFi interference turned on and off. We see negligible variation in the relationship between PRP and distance (~\Cref{fig:benchMark}(right)). This is because the three advertising channels of BLE fall between or outside the main frequencies used for IEEE 802.11, allowing for better coexistence with WiFi.

Does interference from many co-located beacons hurt PRP? BLE beacons send out short advertising messages in passive mode containing a payload of at most 31 bytes. As pointed out in ~\cite{robinBLE}, the small size of the advertising messages helps in avoiding any significant collisions of upto 200 or more co-located devices. Similarly, the co-location of many receivers or scanning devices will not impact PRP. In our set-up, the receiver receives the advertising message in a passive scanning mode, and does not respond in any way. As a result, many scanning devices do not lead to any interference.

% This is because modern wireless schemes have in-built interference avoidance that allows packets to be transmitted in spite of interference.

%We present some micro benchmark results on Packet Reception Probability(PRP). ~\Cref{fig:benchMark} shows the results. We make the following observations---

%\begin{itemize}
	%\item There is no one-to-one relation between PRP and RSSI. We collected both PRP and RSSI data at same locations. We averaged the RSSI values over all the packets used for PRP and then drew a scatter plot between aggregated RSSI and PRP ~\Cref{fig:benchMark} (left). We get large variance in RSSI when PRP is low and high.
	%\item PRP values are robust to the smartphone device variation. We collected PRP values at same location with two models---Nexus5X and NuuA4L. In ~\Cref{fig:benchMark} (middle), we can see minor differences in packet reception which are caused because the experiments with the phones were conducted on different days.
	%\item PRP values are robust to ambient WiFi interference. We conducted two experiments---one where no ambient WiFi was present. In the other case, ambient WiFi was present and we saturated the WiFi using iPerf tool. In ~\Cref{fig:benchMark} (right), we can see that both PRP curves almost converges.
%\end{itemize}

\section{Results}
\label{sec:Results}

We compare the localization performance of baselines against B-PRP in ~\Cref{sub:Accuracy Evaluation}. For these results, we assume that all beacon and reception locations are known (for all methods). In ~\Cref{sub:Robustness}, we evaluate the robustness to the number and placement of beacons. In ~\Cref{sub:Contact Tracing Accuracy Evaluation}, we evaluate the contact tracing performance of the PRP against RSSI.
In ~\Cref{sub:ImpactBeaconSetup} and ~\Cref{sub:ImpactRetrainig}, we list the results for B-PRP when we reduce the beacon set-up costs and the number of labelled training locations. In summary: %Finally, in ~\Cref{sub:ImpactRetrainig} we show the accuracy of B-PRP when retrained inexpensively using traces collected at random locations by store workers. In summary:
\begin{itemize}
  \item Median error for B-PRP is $1.03m$ and $1.45m$ in library and retail store. The corresponding errors for the best baseline, Bayesian RSSI are $1.3m$ and $2.05m$.
	\item Median error for contact tracing distance estimation with PRP is $0.97m$ and $1.22m$ in library and retail store. The corresponding errors with RSSI are $1.69m$ and $1.25m$.
  \item B-PRP is more robust than RSSI to decreasing number of beacons. With $5$ beacons, B-PRP performance is $65\%$ better in the library and $50\%$ better in the retail store.
  \item B-PRP performs better than Bayesian RSSI when we use only Non Line-of-Sight(NLOS) or far away beacons. With beacons placed greater than $6m$ distance, B-PRP gives error of $1.53m$ and $2.07m$ in LOS and NLOS. RSSI errors are $3.85m$ and $5.15m$.
	% \item When we fuse B-PRP with RSSI, we see better error $~0.5m$ at small distances of (\SI{\le2}{\meter}). At large distances (\SI[]{\ge 2}{\meter}), errors in RSSI make fusion results to be worse than B-PRP.
  \item B-PRP can reduce set-up cost by learning most beacon locations. Given data from 12 training locations, B-PRP needs to know exact location of only $6$ beacons and it can infer the remaining $54$ beacon locations while giving an accuracy of $1.05m$.
  \item B-PRP can reduce retraining efforts by leveraging data from unknown locations. Having data from 12 known locations vs (6 known + 6 unknown) locations gives the same accuracy level. We can improve accuracy $\sim 40\%$ by adding data from unlabeled spots.
\end{itemize}
% Finally we compare location accuracy against baselines.

\subsection{Localization Accuracy Evaluation}
\label{sub:Accuracy Evaluation}
\begin{table*}[b]
  \centering
  \begin{tabular}{cccccccc}
    \toprule
    Environment & \textbf{B-PRP} & B-PRP + RSSI & Bayesian RSSI~\cite{madigan2005bayesian} & Horus~\cite{youssef2005horus} & Bayesian FP~\cite{chen2013bayesian} & MCL~\cite{hu2004localization}\\
    \midrule
    Library & 1.03m & 0.91m $(\downarrow11.6\%)$ & 1.3m $(\uparrow26.2\%)$ & 1.83m $(\uparrow77.6\%)$ & 1.93m $(\uparrow87.4\%)$ & 2.26m $(\uparrow119\%)$\\
    Retail Store &  1.45m & 1.46m $(\uparrow0.6\%)$ & 2.05m$(\uparrow41.4\%)$ & 1.85m$(\uparrow27.6\%)$ & 1.95m$(\uparrow34.5\%)$ & 2.93m $(\uparrow102\%)$\\
  \bottomrule
\end{tabular}
% \vspace{-0.1in}
\caption{\textbf{Median error (in $m$) of B-PRP and baselines:} B-PRP perfoms best in both environments followed by Bayesian-RSSI in the library and Horus in retail store. Fusion of B-PRP and RSSI performs slightly better in the ideal library environment with many beacons at close distance. Horus and Bayesian FP underperform as they require more training states for better accuracy. All methods perfom worst in the harsh retail environment.
\vspace{-0.25in}
}
\label{tab:errTable}
\end{table*}
\begin{figure}[htb]
    \vspace{-0.1in}
    \centering
    \includegraphics[viewport=0.9in 0.01in 13.1in 5.9in,width=\columnwidth, clip=true]{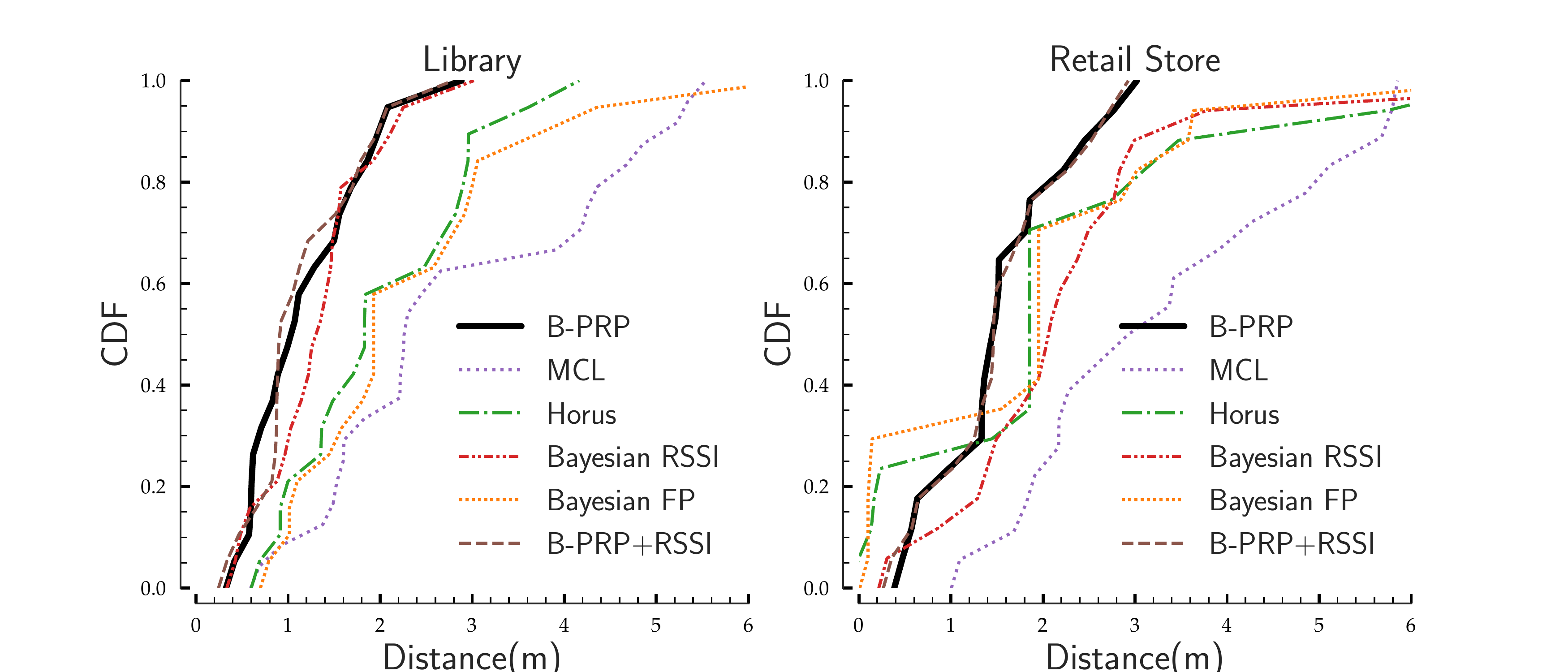}
    \vspace{-0.2in}
    \caption{
    % CDF distribution comparing tracking errors of Bayesian PRP with baselines for library and retail store. Please note that for RSSI techniques, we have averaged RSSI value across the same number of packets that was used by PRP technique. So both PRP and RSSI use the same amount of information.
    \textbf{CDF error  distribution} for Bayesian PRP and baselines in library and retail store. For RSSI techniques, we have averaged RSSI value across the same number of packets that was used by PRP technique.
    %RSSI results are not impacted by the variance in RSSI value of a single packet.
    % One interesting observation for the retail environment---error of fingerprinting techniques like Horus and Bayesian FP are initially lower which corresponds to the cases where the testing and training locations coincided. The errors dramatically increase for all other cases.
    }
    \vspace{-0.1in}
    \label{fig:imageResults}
\end{figure}

We compare the accuracy of B-PRP against baselines. We use Euclidean distance to measure the error between actual and estimated locations for each time window. We show cumulative distribution over errors in ~\Cref{fig:imageResults} and median error in ~\Cref{tab:errTable}.

% \begin{figure}[htb]
%   \centering
%   \includegraphics[viewport=0.9in 0.03in 13.1in 5.9in,width=\columnwidth, clip=true]{losResults_ver2}
%   \caption{ \textbf{CDF of localization errors} for B-PRP and RSSI for different scenarios. In library, B-PRP performs much better when asked to localize with \textit{only} non-line-of-sight beacons.
%   \vspace{-0.2in}}
%   \label{fig:imageLoSResults}
% \end{figure}

First, observe that B-PRP achieves a median error of $1.03m$ and $1.45m$ in the library and retail store. The next best method, Bayesian RSSI, achieves errors of $1.3m$ and $2.05m$. The errors for all methods are higher for the retail store which has more human traffic than the library. B-PRP can outperform baselines due to two reasons: (a) B-PRP can extract information even from lost packets, and (b) It incorporates a new multipath-model that can work in the presence of obstacles. The stack model helps to increase the median accuracy of B-PRP from $1.41m$ to $1.03m$ in the library and from $1.6m$ to $1.45m$ in the retail store. B-PRP performs much better than RSSI with non-line-of-sight (NLoS) beacons. The median error for RSSI is $2.34m$ with NLoS beacons in the library compared to $1.63m$ with B-PRP.

% \begin{figure}[htb]
%   \vspace{-0.1in}
%   \centering
%   \includegraphics[viewport=0.9in 0.01in 13.1in 5.9in,width=\columnwidth, clip=true]{losResults_ver2}
%   \vspace{-0.2in}
%   \caption{ \textbf{CDF of localization errors} for B-PRP and Bayesian RSSI for different scenarios. In library, B-PRP performs much better when asked to localize with \textit{only} non-line-of-sight beacons.
%   \vspace{-0.2in}
%   }
%   \label{fig:imageLoSResults}
% \end{figure}

% \vspace{4pt}\noindent\textbf{Non-Line of Sight: }How does Bayesian RSSI and B-PRP compare in line-of-sight (LoS) vs non-line-of-sight (NLoS) scenarios? In ~\Cref{fig:imageLoSResults}, we can see that the median error for Bayesian RSSI is $2.34m$ with NLoS beacons in the library, which is far more than $1.3m$ error that it achieves using all beacons. B-PRP achieves a better median error of $1.63m$ with NLoS devices. This shows an additional advantage of B-PRP over traditional approaches using RSSI. It can perform significantly better even when no beacon is in line-of-sight. Note that this error difference is significant because an additional error of $0.8m$ can significantly worsen false-positive and false-negative rates in end user applications.

\begin{figure}[htb]
  \centering
  \includegraphics[viewport=0.8in 0.01in 13.1in 5.9in,width=\columnwidth, clip=true]{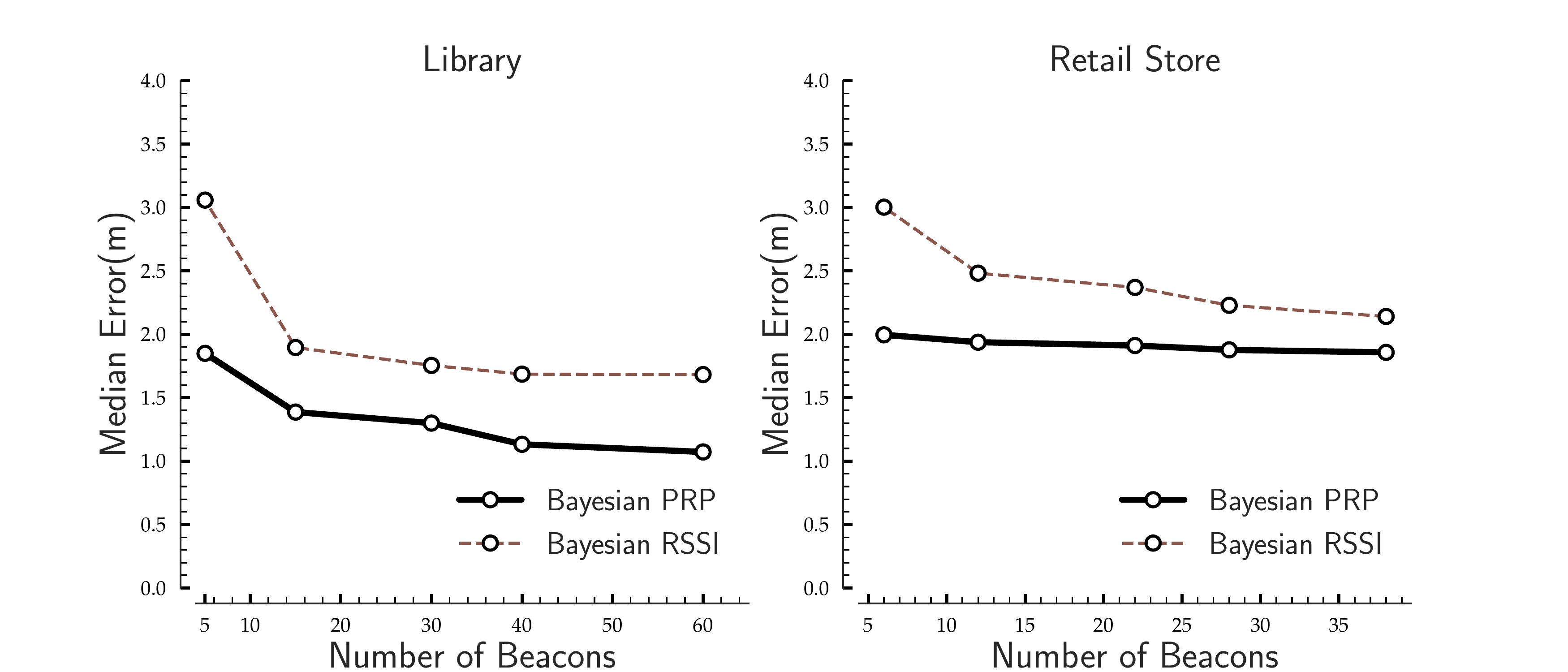}
  \caption{\textbf{Variation in median error for B-PRP with beacon number}. The error is within $2m$ for all cases. With $5$ beacons, B-PRP performance is better than Bayesian RSSI: $65\%$ (library) and $50\%$ (retail store).
  \vspace{-0.2in}}
  \label{fig:robustnessComparisonBeacon}
\end{figure}

\vspace{4pt}\noindent\textbf{B-PRP+RSSI: } One might wonder if B-PRP can be augmented with RSSI to achieve even better performance. We augment B-PRP with RSSI to test this hypothesis. As shown in ~\Cref{fig:imageResults}, the method works approximately similar to B-PRP. As we demonstrate in the next subsection, this is because at smaller distances, RSSI experiences little packet loss and helps our model make better inference. However, at large distances, RSSI experiences larger sampling bias and consequentially, just acts as noise, thereby hurting the model.

\subsection{Beacon Number and Placement}
\label{sub:Robustness}

We evaluate the robustness of B-PRP against the best performing baseline---Bayesian-RSSI to two factors---the number of beacons and the placement of beacons.

% Here, we conduct experiments to evaluate the robustness of the Bayesian Packet Reception Probability. We test robustness against two factors---the number of beacons and the placement of beacons.
% Notice that this is different from the case where we want to reduce set-up cost by making the beacon locations unknown. Here we want to see how the performance of B-PRP varies if we reduce the total number of beacons. Also, how does the placement or distance of a beacon with respect to the receiver affect accuracy. We choose the best performing baseline---Bayesian-RSSI for comparison.
\textbf{Beacon Number:}
% In our test beds, we initially set up the beacons at a distance of $1m$. That resulted in 60 beacons in total in the library and 38 beacons in the retail store. A large retail store will need to place hundreds of beacons to maintain this inter-beacon distance, which in turn increases the localization infrastructure and configuration cost. So, it's natural to ask :\textit{how do we impact localization accuracy when we decrease the number of beacons?}.
We evaluated the accuracy with fewer number of beacons (lower bound is set to three beacons -- the minimum required to localize). Lower number of beacons will reduce the localization infrastructure cost
% We re-ran the experiments with a fewer number of beacons. Since we need at least three beacons to localize uniquely, we use three as the lower bound for the number of beacons. For each number of beacons, we estimate the locations with B-PRP and Bayesian-RSSI.
% \begin{figure}[htb]
% % \vspace{-0.15in}
%   \centering
%   \includegraphics[viewport=0.8in 0.01in 13.1in 5.9in,width=\columnwidth, clip=true]{RobustnessComparisonBeacon_ver3}
%   \vspace{-0.2in}
%   \caption{ \textbf{Variation in median error for B-PRP with beacon number}. The error is within $2m$ for all cases. With $5$ beacons, B-PRP performance is better than Bayesian RSSI: $65\%$ (library) and $50\%$ (retail store).
%   \vspace{-0.2in}
%   }
%   \label{fig:robustnessComparisonBeacon}
% \end{figure}
In ~\Cref{fig:robustnessComparisonBeacon}, we see B-PRP performance degrades slowly than Bayesian-RSSI to decreasing beacon density. The median error of localization for B-PRP is always within $2m$. For Bayesian-RSSI, with lower beacons, the error is as high as $3m$. With $5$ beacons, B-PRP performance is $65\%$ better than Bayesian-RSSI in the library and $50\%$ better in the retail store. Also, note that just with $5$ beacons, B-PRP performs better or equal to Bayesian-RSSI with upto $60$ beacons. This, yet again, demonstrates that the errors in RSSI-based positioning cannot be solved by just additional deployments, but are fundamental (sampling bias and multipath).
%\textit{With limited beacons, B-PRP gives better localization.}
% In ~\Cref{fig:robustnessComparisonBeacon}, we can see B-PRP is far more robust than Bayesian-RSSI for the store. The median error of localization for B-PRP is always within $2m$. For Bayesian-RSSI, with lower beacons, the error is as high as $3m$. For the library, Bayesian-RSSI performance degrades slowly when we have more beacons. But when the number of devices is lower than $10$, B-PRP performs significantly better. With $5$ beacons, B-PRP performance is $65\%$ better than Bayesian-RSSI in the library and $50\%$ better in the retail store. B-PRP performance degrades slowly than Bayesian-RSSI with decreasing beacon density. \textit{When we have limited beacons, we should rely on B-PRP for better and robust location estimates.}

\begin{table*}[b]
  \centering
	\begin{tabular}{cccccccccccc}
		\toprule
		\multirow{ 2}{*}{\shortstack{Line-Of-Sight \\ Condition}} &  \multicolumn{3}{c}{distance $<2m$} & \phantom{abc} & \multicolumn{3}{c}{$2m<$ distance $<6m$} & \phantom{abc} & \multicolumn{3}{c}{distance $>6m$}\\
		\cmidrule{2-4} \cmidrule{6-8} \cmidrule{10-12}
		& B-PRP & RSSI & B-PRP + RSSI && B-PRP & RSSI & B-PRP +RSSI && B-PRP & RSSI & B-PRP +RSSI\\
		\midrule
		LOS & 0.89m & 0.5m & 0.5m && 0.63m & 3.57m & 0.62m && 1.53m & 3.85m & 1.56m\\
		Non-LOS &  0.85m & 1.05m & 0.57m && 5m & 5.95m & 5.62m && 2.07m & 5.15m & 2.7m\\
	    \bottomrule
    \end{tabular}
\caption{
\textbf{Robustness To Beacon Placement:} Recall, our median error using all beacons is $1.03m$. With only beacons that are closer than $2m$ to the receiver, both PRP and RSSI errors are low. Fusion of PRP and RSSI gives even lower errors of $0.5m$. In this range, RSSI values have less variance and more distance information. With beacons further than $2m$, RSSI variances increase which cause fusion results to be worse.}
% PRP gives high error if beacons are between $2m$ and $6m$ in a Non Line-of-Sight scenario. This is expected because the variance in the PRP estimates are high in this region. PRP again gives low error if beacons are further than $6m$. }
\vspace{-0.2in}
\label{tab:PRPvsRSSItable}
\end{table*}

\textbf{Beacon Placement:} \textit{How does the placement of a beacon with respect to the receiver impact the localization accuracy by B-PRP and RSSI?} If we use only beacons that are closer than $2m$ to the reception location, both PRP and RSSI errors are good (c.f.~\Cref{tab:PRPvsRSSItable}). RSSI performs slightly better in Line-of-Sight scenario due to the less variance in RSSI values and more distance information at very close range. Fusion of B-PRP and RSSI also yields lower errors. When beacon distances become greater than $2m$, RSSI errors dramatically increase due to variance in RSSI values caused by multi-path and sampling bias. In comparison, PRP errors are much lower in order of $1.53m$ and $2.07m$ when beacons are more than $6m$ away from the receiver. Errors in RSSI also cause fusion results to be worse. Error for all approaches is high when we use only beacons, all of which are in a Non Line-of-Sight(NLOS) scenario and are at a distance between $2m$ and $6m$ from the receiver. This experiment highlights the importance of PRP. As RSSI estimates suffer from higher sampling bias with increasing distance, the underlying location information gets corrupted. This is why at larger distances, both Bayesian RSSI and B-PRP+RSSI do worse.

\subsection{Evaluating Contact Tracing Distance Estimates}
\label{sub:Contact Tracing Accuracy Evaluation}

\begin{figure}[htb]
    \vspace{-0.1in}
    \centering
    \includegraphics[viewport=0.9in 0.01in 13.1in 5.9in,width=\columnwidth, clip=true]{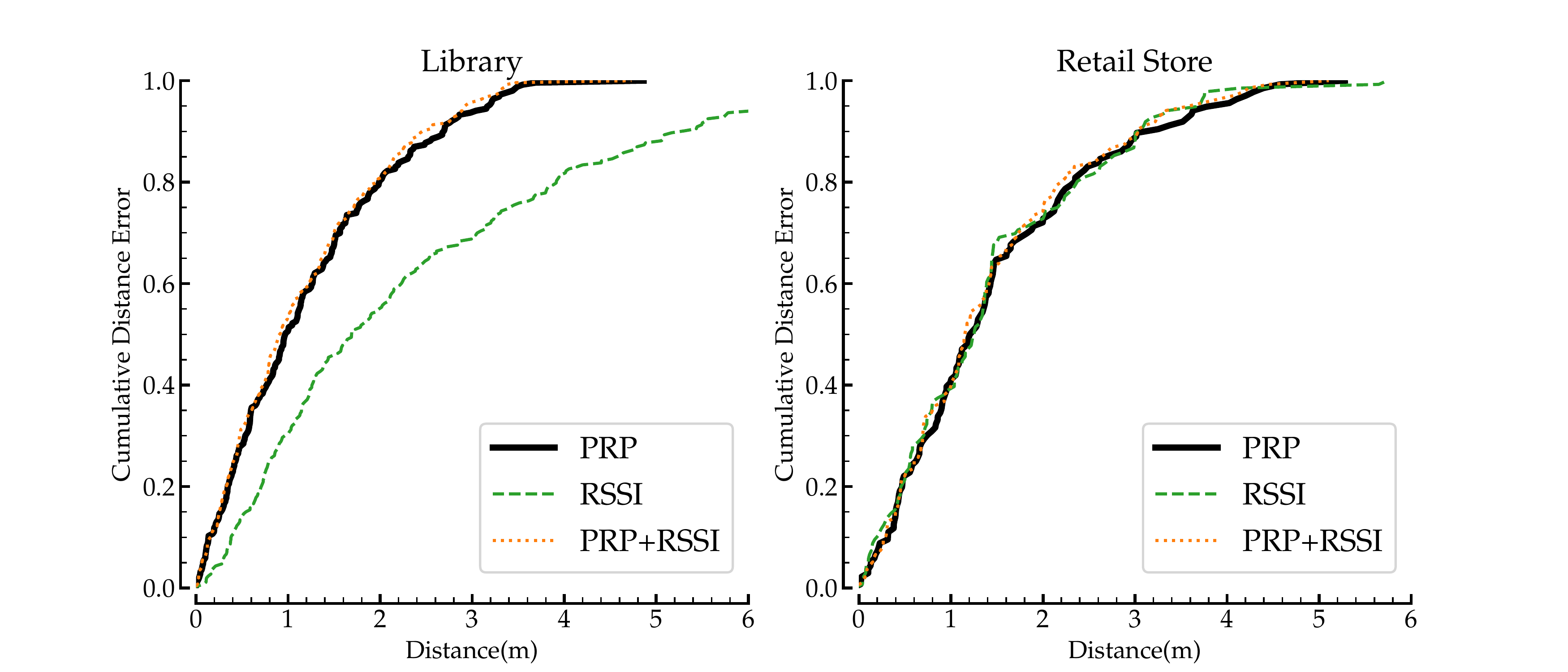}
    \vspace{-0.2in}
    \caption{
    \textbf{CDF error  distribution of contact tracing distance} for PRP, RSSI and PRP+RSSI in library and retail store. PRP and PRP+RSSI gives the best median error in both environments.
    }
    \vspace{-0.1in}
    \label{fig:contactTracingResults}
\end{figure}

We compare the accuracy of PRP against RSSI and PRP+RSSI in contact tracing distance estimation. We measure the absolute error between actual and estimated distances for each pair of person or receivers. We show cumulative distribution over errors in ~\Cref{fig:contactTracingResults}.

PRP achieves a median distance error of $0.97m$ and $1.22m$ in the library and retail store. RSSI achieves distance errors of $1.69m$ and $1.25m$. PRP+RSSI performs the best with median distance errors of $0.91m$ and $1.15m$.

% Now, notice that for a contact tracing application, errors in all distance estimations $d_{i,j}$ between a pair of individuals $i,j$ are not equally important. When the actual distance is  $d_{i,j} < 2m$  (the social distancing range), the errors become more important. When two individuals are far away from $d_{i,j}  \gg 2m$, errors become less important.
% We used a standard risk function described in \cite{van2020aerosol} to map the distance values to a risk metric:
% $$risk = \frac{1.0}{1 + e^{9.17*(d_{i,j}-2.0)}}, $$, where, $d_{i,j}$ is the distance between a pair of individuals $i,j$.
%
% Notice that the risk is a sigmoid function that takes high values $risk \approx 1$ when  $d_{i,j}< 2m$, and low values $risk \approx 0$ when $d_{i,j} \geq  2m$ . We calculate the absolute error between the actual risk (based on ground truth distance) and predicted risk (based on the estimated distance). PRP gives a median risk error of $5 \times 10^{-6}$ and $0.16 \times 10^{-6}$ in the library and retail store, while RSSI achieves $4 \times 10^{-3}$ and $0.22 \times 10^{-6}$ respectively. Note that the median risk error with RSSI in the library is 1000 times larger compared to PRP. PRP+RSSI has risk error of $1.3 \times 10^{-6}$ and $0.06 \times 10^{-6}$. Also, though the risk errors look small, at population scale and across multiple interactions, the risk is very high.

\subsection{Minimizing beacon set-up cost}
\label{sub:ImpactBeaconSetup}

% \begin{table*}[htbp]
%   \centering
%   \begin{tabular}{lllll}
%     \toprule
%     $R$ & $b=60$ & $b=6$ & $b=3$ & $b=1$\\
%     \midrule
%     12 & 1.03 & 1.05 $(\uparrow1.9\%)$ & 1.24 $(\uparrow20.4\%)$ & 1.38 $(\uparrow33.9\%)$ \\
%     8 & 1.05 & 1.22 $(\uparrow16.2\%)$ & 1.82 $(\uparrow73.3\%)$  & 2.15 $(\uparrow104.7\%)$ \\
%     4 & 1.05 & 2.88 $(\uparrow174.2\%)$ & 3.74 $(\uparrow256.1\%)$  & 3.48 $(\uparrow231.4\%)$\\
%   \bottomrule
% \end{tabular}
% \caption{Median error in meters of B-PRP with varying values of $b$ and $R$. $b$ is the number of known beacon locations and $R$ is the number of training locations.
% In each row, we compare performance against $b=60$ which is our baseline. Error increases as we decrease $b$ from left to right in each row. For $R=12$, performance is almost same as with $b=60$ and $b=6$. The error goes up at a higher rate as we move down each column. Effect of decreasing $R$ is more severe than $b$.
% % We can see that the effect of $R$ is more severe compared to $b$.
% }
% \label{tab:autoconfigTable}
% \end{table*}

\begin{table}[htbp]
  \centering
  \begin{tabular}{lllll}
    \toprule
    $N_R$ & $b=60$ & $b=6$ & $b=3$ & $b=1$\\
    \midrule
    12 & 1.03 & 1.05 & 1.24 & 1.38 \\
    8 & 1.05 & 1.22 & 1.82  & 2.15 \\
    4 & 1.05 & 2.88 & 3.74  & 3.48 \\
  \bottomrule
\end{tabular}
% \vspace{-0.1in}
\caption{B-PRP's median localization error (in $m$) with varying number of known beacon locations $b$,  and number of training locations $N_R$ . %In each row, we compare performance against $b=60$ (baseline).
Error increases as we decrease $b$ (each row) and decrease $N_R$ (each column). For $N_R=12$, performance is almost same as with $b=60$ and $b=6$.
% The error goes up at a higher rate as we move down each column.
Decreasing $N_R$ impacts accuracy more than does $b$.
\vspace{-0.1in}
% We can see that the effect of $R$ is more severe compared to $b$.
}
% \vspace{-0.2in}
\label{tab:autoconfigTable}
\end{table}

% \begin{figure*}[t] %!htbp
%   \vspace{-0.15in}
%     \centering
%     \includegraphics[viewport=1.75in 0.0001in 19.1in 4.9in,width=2\columnwidth,clip=true]{autoConfigurationFig1alt_ver1}
%     \vspace{-0.2in}
%     \caption{ \textbf{Reducing beacon set-up cost:} CDF for comparing localization errors of B-PRP as we vary the number of known beacon locations(or primary beacons). As we decrease the number of known beacon locations by order of magnitude from 60 to 6, we hardly see any increase in error when we have 12 training locations. If we reduce training locations to $4$, we need more primary beacons to retain same accuracy.
%     \vspace{-0.1in}
%     % The error increases when we reduce known beacons further to
%     % 1 or we collect training data at less spots, $4$. Reducing training spots has a higher effect on accuracy than decreasing known beacon locations.
%     }
%     % As we decrease the number of master beacons with known
%     % location, the error increases.
%     %
%     % We check the impact of number of training locations on accuracy. We can see that as we
%     % reduce training locations, we need more master beacons to keep accuracy at the level of manually configured framework.}
%     \label{fig:imageAutoConfig1}
% \end{figure*}

% \vspace{-0.15in}
So far, we have used location information of all $B$ beacons. \textbf{Now, we will use the location information for only $b \ll B$ beacons.} $B$ is total number of beacons and $b$ is the number of beacons with known location information. We use data to estimate $B-b$ unknown beacon locations. We then use these estimated values to track a receiver.
% The results shown till now use the exact ground-truth location of all $B$ beacons. Now, we will use the location information for only $b$ master beacons, and estimate the location for the rest $B-b$ beacons within our Bayesian framework. To clarify, for this section $B$ refers to the total number of beacons, $b$ refers to the number of beacons with known locations.
% We will train the model parameters and the unknown beacon locations using data from $R$ training locations. We will then use the trained model parameters for localization. In such a scenario, we want to see what happens to the localization accuracy as we vary the value of $b$. Specifically, we want to reduce $b$ close to 0 and see if we can retain same accuracy as $b=B$ i.e. all beacon locations are known.

We vary the number of known beacon locations $b = \{ 1,3,6,60 \}$. $b=60$ corresponds to when we know all beacon locations. We also vary the value of $N_R$ i.e. the total number of training locations. Ideally, we would like to have less known beacons $b$ and less training locations $N_R$.
% Since we make the beacon locations an unknown parameter in our framework, we also want to get a sense of the number of training locations required to efficiently estimate all these unknowns and retain accuracy level. Ideally, we would not like to use a lot of training locations to compensate for the unknown beacon location, as this adds to the set-up cost.
We show the results in ~\Cref{tab:autoconfigTable}.
% \Cref{fig:imageAutoConfig1}

% We show the results for three different values of $N=\{12, 8, 4\}$. That is, the figure shows the effects of training the model for different values of $k = \{1,3,6,60 \}$, and the variation in error, when we reduce the number of training locations. As a reminder, our manual configuration case is with ($N=12, k=60$), when we listen at ($N=12$) spots with all beacon locations known ($k=60$).
We highlight three observations. \textit{First}, when ($N_R=\{12, 8\})$ there is negligible difference in the CDF of the tracking errors between the cases of $b=60$ and $b=6$. \textit{Second}, for any value of $N_R$, the errors increase when we decrease $b$, with the effects most pronounced for $N_R=4$. \textit{Finally}, the figures suggest that the effect of unknown beacon locations is \textit{less significant} than the effect of the number of training locations. B-PRP can give the same level of performance with as low as $b = 3$ primary beacons when the number of training locations $N_R$ is high. If we reduce $N_R$ to 8, we need at least $b=6$ known beacons.

These results highlight that B-PRP can be deployed for public spaces with little overhead. A retail store operator can just place beacons at random locations, and move around with a smartphone to some known locations. B-PRP can infer the beacon location on its own (for most beacons) and still achieve competitive performance.
% With decreasing $R$, the requirement for $b$ increases more steeply.
%This suggests that B-PRP can reduce the number of known beacon locations close to zero if we have enough training locations. This is intuitive because: every beacon adds new information to $N_R$ training locations, whereas every new training spot allows us to make inferences about all $B$ beacons; since typically $B \gg N_R$, removing one training location will have more impact than one more unknown beacon location.
%Thus, \textit{B-PRP can infer most beacon locations from data.}

\begin{figure*}[t] %!t
  %\vspace{-0.15in}
	\centering
    \includegraphics[viewport={1.75in 0.0001in 19.1in 4.9in},width=2\columnwidth,clip=true]{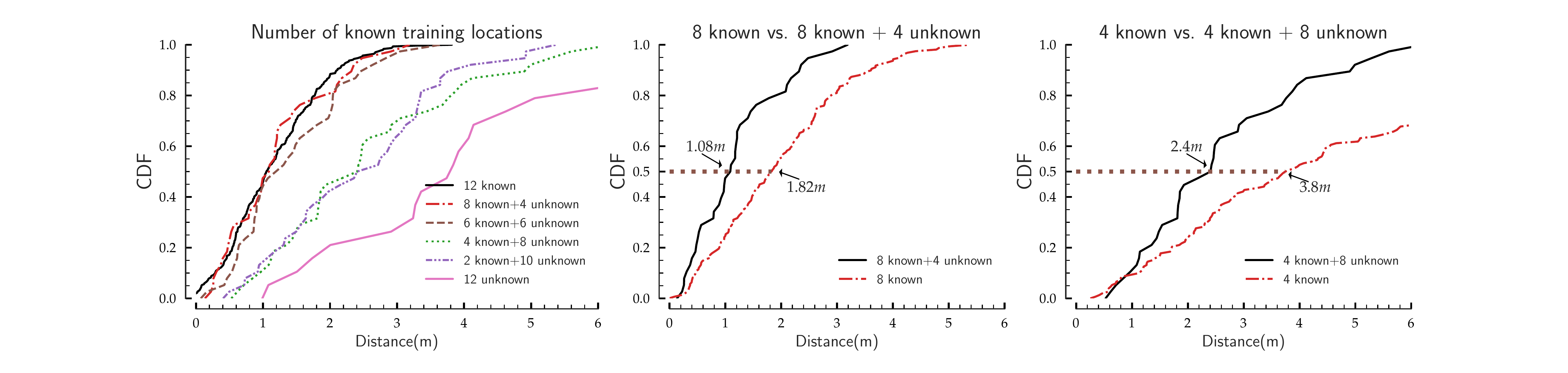}

  \vspace{-0.2in}
	\caption{ \textbf{Reducing retraining efforts:} CDF for comparing errors of B-PRP when we train using data from some known and mostly unknown locations.
  % As we collect data at more unknown locations than known, errors increase.
  If we have data from 12 known locations vs (6 known + 6 unknown) locations, we get the same accuracy level.
  % This means that the store workers can collect data from 12 locations, but now they need to annotate only half of those and B-PRP ensures same accuracy level.
  % The right two subfigures go on to show that collecting data at unknown locations is actually valuable. In both cases, we improved accuracy $\sim 40\%$ by adding data from unknown spots rather than only using data from known spots.
  In the right two subfigures, we show that we improved accuracy $\sim 40\%$ by adding data from unknown spots rather than only using data from known spots.}
  \vspace{-0.1in}
  % Median errors remain the same when we cut the number of known training spots to 6 which is half of what we originally had.
  % The two subfigures on the right show that it is better to add data from unknown spots than collecting data at fewer locations. Median errors improve by $\sim 40\%$}
  % We check the impact of number of known training locations on accuracy of configured framework. Here we collect data at the same number of spots, but we know the location information for less number of spots
	% as we move from left to right. We can see that we need more master beacons to compensate for the unknown training spots.}
	\label{fig:imageAutoConfig2}
\end{figure*}

\subsection{Reducing training efforts}
\label{sub:ImpactRetrainig}

Till now, we have used the location information of all training spots $N_R$ while training. Now, \textbf{let's use the information for only $r < N_R$ training spots} and estimate the remaining $N_R-r$ locations using our framework.
% The B-PRP model used for location estimation till now was trained using data from $R=12$ known locations. That is we stood at $12$ carefully marked spots and collected data. Now, we will use the location information for only $r$ training locations and consider the rest $R-r$ as random or unknown spots, which we will estimate within our framework. $R$ is the total number of training locations. We estimate the model parameters and unknown training locations using this data. Then we apply the trained model parameters for localization. In such a scenario, we want to see what happens to the localization accuracy as we vary the value of $r$.

We change the value of known training locations $r =\{ 12, 8, 6, 4, 2, 0\}$, with $N_R=12$. \Cref{fig:imageAutoConfig2} shows the results. In the leftmost sub-figure, we see that as $r$ decreases, error increases; but notice that we can cut the known locations in half, from $r=12$ to $r=6$, without appreciable increase in error. This means that we can collect data from 12 spots but need to annotate only half of those and B-PRP can still maintain the same accuracy level. One might wonder, \textit{do we really gain any performance improvement by adding data from unknown locations?} \Cref{fig:imageAutoConfig2} (two right sub-figures) validate that conjecture. Suppose, our training dataset contains data from $12$ training locations in total. Now, $8$ of those are labeled with location information while $4$ are unlabeled. If we train PRP parameters using only $8$ labeled data locations, our median error from the trained model is $1.82m$. In contrast, if we use the entire dataset and treat the location of the $4$ unlabeled data points as random variables in our framework, we improve the median error to $1.08m$. Similarly, if we have $4$ labeled and $8$ unlabeled locations, by using all the locations our errors improve from $3.8m$ to $2.4m$. Thus, data from un-labled locations are valuable for training PRP parameters. This further eases the deployment cost by allowing operators to collect fewer labelled data points.

%An implication for a large retail store---\textit{PRP model can be seamlessly retrained using data from random locations and we need to annotate only a small fraction of those locations.}

%!TEX root = main.tex
\vspace{-0.1in}
\section{Discussion and Limitations}
\label{sec:Discussion}

A few points are worth noting:

\noindent\textbf{Applicability to general indoor environments: } We design B-PRP with a focus on public indoor environments like retail stores that have stacked layouts. This layout is applicable to multiple spaces like libraries, warehouses, pharmacies, etc. and covers an important application area. While the current multipath-resilience model of B-PRP does not directly apply to other environments like homes, we believe PRP itself is applicable to such environments and provides the unique advantage of robustness at large distances. Furthermore, in such environments, obstacles like walls can be modelled using the approach followed in B-PRP.

\noindent\textbf{Access to Layouts: }We design the layout requirement for B-PRP to be low-effort. The layout and stacks can simply be extracted from the floorplan of the store, either manually or through an app. This makes the deployment effort low. Furthermore, B-PRP can apply to store layouts with more stacks than the ones used in this paper. We may encounter geometric elements like \textit{three stacks away} ($3-S$), \textit{four stacks away} ($4-S$) etc. We do not necessarily need a separate PRP function for each of these elements. Since PRP becomes very low after certain number of stacks, we can club these spaces into one geometric element and learn a single model.

 \noindent\textbf{Computational complexity:}  Bayesian MCMC techniques may take more time to infer location. We ran our computations in python on a MacBook Pro laptop with $2.5 GHz$ Intel Core i7 processor and $16GB$ RAM. With $60$ beacons, it took us $\sim 3$ seconds to find the next location, within our time resolution ($\delta=10$s) for localization. We can further speed-up by using native code and parallelizing the inference.
% used in our experiments. Please note that the process can be further sped-up by parallelizing the code or writing it in C.
% \noindent \textbf{Additional infrastructure:} Bayesian PC uses commercially available Bluetooth Low Energy (BLE) beacons as added infrastructure for localization.
% % Our framework is agnostic to the type of device. We can easily
% % apply the techniques to packets received from WiFi access points.
% We chose to work with beacons as they are relatively inexpensive and are being rapidly deployed over retail industry to enable proximity marketing. Today we can buy a beacon as low as USD $5\$$. According to report by \cite{ProximityDirectory},
% $75\%$ of U.S. retailers including Walmart, Walgreens, Apple Retail, Home Depot, Target etc. have already added beacons to their marketing mix. \cite{ABIresearch} forecasts that more than 500 million BLE units will be installed by 2021 causing a
% $133\%$ compound annual growth rate in BLE retail market. We say that the BLE technology will become commonplace as WiFi APs in retail stores and will not be treated as added infrastructure in near future.

 \noindent\textbf{Scalability to the number of packets:} One limitation of B-PRP is that it needs more than one packet to localize.
%  It needs multiple packets to calculate the packet count value. That makes it less scalable in terms of the number of packets.
We can reduce the number of packets used for localization by changing the advertising frequency. We observe in our experiments that as we lower the sending rate from $10Hz$ to $1Hz$, while keeping the localization rate to once per 10 seconds, the median error increases by just $0.2m$. %One option: combine our framework with Bayesian RSSI to make it more packet scalable.

% \textbf{Applicability to very crowded stores:} Bayesian PRP performance degrades in the presence of many people in the localization environment. We experimented once in the retail store during its peak hour of customer visit. At least 15 (max 20) people were always moving around in the retail store (10m by 10m).
% Performance of both PRP and RSSI techniques became significantly worse in this adverse scenario.
% % The median error of Bayesian PRP in these adverse scenarios became $\sim 2.5m$.
% Currently, we are exploring a peer-based PRP system like \cite{liu2012push} that leverages the crowd presence to boost its performance. %We always had at least five customers moving around in the same space. Additionally, we also had store workers who were organizing items in the stacks at that time. The median error of Bayesian PRP in these adverse scenarios became $\sim 2.5m$. Currently, we are exploring a peer-based PRP system like \cite{liu2012push} that leverages the crowd presence to boost its performance.
% % and thus achieves robust localization in presence of humans.

%!TEX root = main.tex
\vspace{-0.15in}
\section{Related Work}
\label{sec:Related Work}

%We provide a brief overview of the related work in indoor localization.
We can classify localization art on different factors--- communication signal used for localization, models to relate distance and signal properties. Most works use signals exchanged with anchor nodes(known location) to infer location of target.
% Anchor nodes are devices that take part in radio communication with the nodes which have unknown location.
Anchor nodes can be ---WiFi access points \cite{bahl2000radar}, Bluetooth beacons \cite{zhao2014does,wu2017ibill,montanari2017study}, FM radios \cite{chen2012fm}, Zigbee devices\cite{lau2009measurement}, ultra-wide band(UWB) devices  \cite{grosswindhager2018salma}, RFID tags \cite{wang2013dude,yang2015accurate,jiang2018orientation,DBLP:conf/ipsn/Jiang0YGL19}, ultra-sound emitters \cite{huang2014shake}, light emitters \cite{liu2017smartlight,kuo2014luxapose,zhang2017pulsar,zhu2017enabling},60GHz devices \cite{DBLP:journals/twc/PalaciosBCW19,DBLP:conf/infocom/BielsaPLSCW18}, sub-centimeter sized devices \cite{SensysNandakumarIG18}. In contrast, we use BLE beacons which offer advantages over the others. WiFi access points and cameras require continuous power and are more expensive than BLE beacons, which run on long-lasting batteries (lasting $3$ to $5$ years). A store can deploy hundreds of BLE beacons at a lower cost than WiFi access points or video cameras. We can scale BLE-based systems through past work in opportunistic listening that ensures better channel sharing \cite{robinBLE}. WiFi, while widely available in public spaces such as malls and coffee shops, are often absent in large indoor retail stores (e.g., Walmart), in part because the presence of WiFi allows individuals in the store to comparison shop, putting the physical store at a competitive disadvantage. While \cite{grosswindhager2018salma} shows the promise of low-cost UWB sensing, the solution requires the widespread adoption of UWB tags to track objects. With BLE, we can track consumers via their Bluetooth enabled smartphones.

The localization techniques use different signal property
--- RSS or received signal strength\cite{bahl2000radar,chintalapudi2010indoor,yang2009indoor}, CSI or channel state information \cite{tian2018performance,wang2014eyes},
AoA or angle of arrival \cite{xiong2010secureangle,xiong2013arraytrack,kumar2014accurate} , ToF or time-of-flight \cite{mariakakis2014sail,sen2013avoiding,sen2015bringing}. AoA, ToF and CSI systems require hardware level changes on the receiver side and thus cannot be used by a retail store with customers who use commodity smartphones.
Range free techniques use less accurate proximity information \cite{hu2004localization,rudafshani2007localization,he2003range}.
% They cannot provide accurate location due to lack of ranging information \cite{zhao2013localization}.
We use a new property---packet reception probability which is light weight and can be easily deployed on commercial smartphones.

Received Signal Strength (RSSI) systems are broadly of two types---model-based and fingerprint-based. Model-based techniques \cite{madigan2005bayesian,chintalapudi2010indoor,banerjee2010virtual} represent RSSI loss between anchor and target as a function of distance.
% During tracking, they use the function to find the distance from RSSI values and apply trilateration to infer location.
Fingerprint-based techniques \cite{bahl2000radar}, \cite{youssef2005horus} build a map of probable RSS values
from anchor nodes at sampled locations.
% \citet{wen2015fundamental} provide a theoretical study on the accuracy limits of fingerprinting techniques.
% Building the fingerprint map is a huge training overhead. \cite{rai2012zee,park2010growing,yang2013freeloc,wu2013will} reduce the overhead via crowdsourcing. \citet{chen2016graph} use a graph-based method for easy fingerprinting.
% In this paper we show a Bayesian MCMC based approach for easy calibration of our framework.
% Training workloads are much lower for our method which are further reduced by a easy-to-configure framework.
% Most RSSI based techniques have high errors due to multi-path effects, non line-of-sight, fading, human interference, etc. \citet{tian2018improve} propose a theoretical improvement in accuracy by studying temporal correlation of RSS. \cite{guo2010perpendicular} indirectly uses RSSI values and the geometric relationship of a perpendicular intersection to localize.
Here we use a more robust property and design an easy-to-configure framework.

In this paper, we study tracking for public spaces like retail stores which have attracted attention due to proximity marketing \cite{ProximityDirectory}.
% states more than $75\%$ of U.S. retailers are inclined towards smart stores.
% All the above techniques for such spaces suffer from reliability, scalability or deployment issues.
% and activity inference inside a store is a challenging problem which the above methods fail to solve. They either
\citet{radhakrishnan2016iris,radhakrishnan2016iot+} look at the problem of inferring item interaction in stores using wearable sensors. iBILL \cite{wu2017ibill} jointly uses iBeacon RSSI model and inertial sensors to localize in supermarkets with $90\%$ error less than $3.5m$. Tagbooth \cite{liu2015tagbooth} , ShopMiner \cite{shangguan2015shopminer} tracks
customer interaction with commodities using RFID tags in retail stores.
% Cupid \cite{sen2015bringing} combines RSS and ToF information to build a scalable tracking system.
% \citet{mao2018scalability} provide a theoritical study on the scalability of WiFi fingerprinting techniques with the number of users.
% Here we solve retail tracking using packet reception probability.
The closest approach to our work is ~\cite{de2017finding} which counts packets to estimate distance. But here, we estimate using packet reception probability (PRP). We show PRP as a robust estimator of distance, and propose a Bayesian framework to estimate distance using PRP.
% and study the impact of device density, training workload, non line-of-sight on accuracy.

%!TEX root = main.tex
\vspace{-0.1in}
\section{Conclusion}
\label{sec:Conclusion}

This paper establishes the feasibility of using Bluetooth Low-Energy (BLE) to provide a robust, scalable indoor localization solution using commodity hardware. Demonstrating the feasibility of BLE based distance estimation technique is particularly important during the current pandemic, where BLE has emerged as key technology for contact-tracing.
%The contact-tracing apps do not rely on RSSI-based solutions since RSSI suffers from sampling bias and multipath effects. Instead, these apps use BLE for presence, an inadequate proxy for distance.
BLE-based distance estimation today relies on either RSSI or just presence, both of which have publicly documented failure modes. We analyze the fundamental underpinnings of these failure modes and demonstrate robust localization through the Bayesian formulation of a new metric---Packet Reception Probability--that \textit{exploits the absence of received packets}. We show significant improvements over the state of the art RSSI methods in two typical public spaces---a retail store and a library. We show that fusing B-PRP with RSSI is beneficial at short distances (\SI[]{\le 2}{\meter}). Beyond \SI[]{\ge 2}{\meter}, fusion is worse than B-PRP, as RSSI based estimates beyond \SI[]{\ge 2}{m} are effectively de-correlated with distance. Our solution does not require any hardware, firmware, or driver-level changes to off-the-shelf devices, and involves minimal deployment and re-training costs. We have developed a triangle inequality based joint likelihood framework that directly estimates contact tracing distance between two individuals rather than estimating their locations first, which gives us 10\% performance improvement. While our solution is the first step toward robust, reliable indoor contact tracing, we are extending our framework for peer-to-peer distance estimations without beacons (i.e. using only smartphones) for outdoor settings.

\printbibliography
% % \vskip 0pt plus -1fil
% \vspace{-0.8in}
% \begin{IEEEbiography}[{\includegraphics[width=1in,height=1in,clip,keepaspectratio]{subham.jpg}}]{Subham De}
%   is a PhD student in Computer Science at the University of Illinois Urbana Champaign.
% \end{IEEEbiography}
% % \vskip 0pt plus -1fil
% \vspace{-1in}
% \begin{IEEEbiography}[{\includegraphics[width=1in,height=1in,clip,keepaspectratio]{deepak.jpg}}]{Deepak Vasisht}
%   is an Assistant Professor in Computer Science at the University of Illinois Urbana Champaign.
% \end{IEEEbiography}
% % \vskip 0pt plus -1fil
% \vspace{-1in}
% \begin{IEEEbiography}[{\includegraphics[width=1in,height=1in,clip,keepaspectratio]{hari.jpg}}]{Hari Sundaram}
%   is an Associate Professor in Computer Science at the University of Illinois Urbana Champaign.
% \end{IEEEbiography}
% % \vskip 0pt plus -1fil
% \vspace{-1in}
% \begin{IEEEbiography}[{\includegraphics[width=1in,height=1in,clip,keepaspectratio]{robin.png}}]{Robin Kravets}
%   is an Associate Professor in Computer Science at the University of Illinois Urbana Champaign.
% \end{IEEEbiography}
\end{document}